\documentclass[12pt,a4paper]{article} 
\usepackage{graphicx}
\newcommand\TT{\rule{0pt}{2.5ex}}        
\newcommand\BB{\rule[-1.0ex]{0pt}{0pt}}  

\newcommand{\be}{\begin{equation}}
\newcommand{\en}{\end{equation}}
\newcommand{\bea}{\begin{eqnarray}}
\newcommand{\ena}{\end{eqnarray}}
\newcommand{\lbl}[1]{\label{eq:#1}}
\newcommand{\lbltab}[1]{\label{tab:#1}}

\newcommand{\lblsec}[1]{\label{sec:#1}}
\newcommand{\rf}[1]{(\ref{eq:#1})}
\newcommand{\Table}[1]{\ref{tab:#1}}
\newcommand{\fig}[1]{\ref{fig:#1}}
\newcommand{\sect}[1]{\ref{sec:#1}}
\newcommand{\braque}[1]{{\langle #1 \rangle}}
\newcommand{\bc}{\begin{center}}
\newcommand{\ec}{\end{center}}
\newcommand{\bt}{\begin{tabular}}
\newcommand{\et}{\end{tabular}}
\newcommand{\ba}{\begin{array}}
\newcommand{\ea}{\end{array}}
\newcommand{\gapprox}{%
\mathrel{%
\setbox0=\hbox{$>$}\raise0.6ex\copy0\kern-\wd0\lower0.65ex\hbox{$\sim$}}}
\newcommand{\lapprox}{%
\mathrel{%
\setbox0=\hbox{$<$}\raise0.6ex\copy0\kern-\wd0\lower0.65ex\hbox{$\sim$}}}
\newcommand{\inleft}{%
\mathrel{%
\setbox0=\hbox{$<$}\copy0\kern-0.5\wd0\lower1.1\ht0\hbox{$\scriptstyle{in}$}}}
\newcommand{\inright}{%
\mathrel{%
\setbox0=\hbox{$>$}\copy0\kern-0.5\wd0\lower1.1\ht0\hbox{$\scriptstyle{in}$}}}
\newcommand{\outleft}{%
\mathrel{%
\setbox0=\hbox{$<$}\copy0\kern-0.5\wd0\lower1.1\ht0\hbox{$\scriptstyle{out}$}}}
\newcommand{\outright}{%
\mathrel{%
\setbox0=\hbox{$>$}\copy0\kern-0.5\wd0\lower1.1\ht0\hbox{$\scriptstyle{out}$}}}
\newcommand{\e}{{\rm e}}
\newcommand{\im}{{\rm Im\,}}
\newcommand{\Kbar}{\bar{K}}
\newcommand{\mpi }{m_\pi}
\newcommand{\mpid}{m_\pi^2}
\newcommand{\mpic}{m_\pi^3}
\newcommand{\fpid}{F_\pi^2}
\newcommand{\mpiq}{m_\pi^4}
\newcommand{\mkd}{m_K^2}
\newcommand{\mbkd}{\bar{m}_K^2}

\date{\today}
\begin{document}
\bc
{\Large\bf MO analysis of the 
high statistics Belle results on $\gamma\gamma\to \pi^+\pi^-,\ 
\pi^0\pi^0$ with chiral constraints}\\[1.5cm]

{\large R. Garc\'\i{}a-Mart\'\i{}n$^{1,2}$ and 
 \large B. Moussallam$^{2}$}\\[1cm]

{\sl  $^{1}$Departamento  de  F\'\i{}sica  Te\'orica II,  Facultad  de
  Ciencias  F\'\i{}sicas, Universidad  Complutense de  Madrid, E-28040
  Madrid, Spain }\\[0.5cm] 

{\sl $^{2}$Groupe de Physique Th\'eorique, Institut de Physique Nucl\'eaire\\
Universit\'e Paris-Sud 11, F-91406 Orsay, France}\\[2cm]
\ec

\begin{abstract}
\noindent
We reconsider Muskhelishvili-Omn\`es (MO) dispersive represen\-tations
of pho\-ton-photon  scattering to two  pions, motivated by the  very high
statistics results recently released by the Belle collaboration for
charged as well as neutral pion pairs and also by
recent  progress  in  the  determination of  the  low-energy  $\pi\pi$
scattering amplitude.
Applicability of this formalism is extended beyond 1 GeV by taking into
account inelasticity due to $K\bar{K}$ . 
A  modified  MO representation  is  derived  which  has the  
advantage that all polynomial ambiguities are 
collected into the subtraction constants and have simple relations to pion
polarizabilities. It is obtained by treating differently
the exactly known QED Born term and the other components of the
left-hand cut.  
These components are approximated by a sum over
resonances. All resonances up to spin two and masses
up to  $\simeq1.3$ GeV are  included. The tensor contributions  to the
left-hand cut are found to be numerically important. 
We  perform  fits   to  the  data  imposing  chiral
constraints, in particular, using a model independent sum rule result on
the $p^6$ chiral coupling $c_{34}$. Such theoretical  constraints are
necessary   because    the   experimental   errors    are   dominantly
systematic. Results on further  $p^6$ couplings  and pion  dipole and
quadrupole polarizabilities are then derived from the fit.
The relevance of
the  new data  for distinguishing  between two  possible  scenarios of
isospin breaking in the $f_0(980)$ region is discussed.
\end{abstract}

\newpage
\section{Introduction:}

Photon-photon  scattering into  two  pions is  a  process which  probes
several aspects of QCD strong dynamics. In particular, as all 
participating particles  are either massless  or light, it  can probe
the low-energy chiral effective theory of QCD. 
This effective  theory has  now been
worked   out  up  to   order  $p^6$~\cite{bce99a,bce99b}.   There  are
indications that at this order, it can represent the exact dynamics to
a very high precision in the two-flavour  expansion. This was shown,
for instance, for  the $\pi\pi$ $S$-wave scattering lengths (see
e.g. the  review~\cite{Colangelo:2009zz}). 
Unfortunately, most  of the
coupling  constants  of the  $p^6$  chiral  Lagrangian  are still
undetermined. The 
$\gamma\gamma\to\pi\pi$  amplitude is of  particular interest  in this
respect because of its strong sensitivity to several of these $p^6$ couplings.
Physically, these couplings are associated with electric and magnetic
dipole and quadrupole polarizabilities of the pion. These are
important observables associated with the structure of the pion. They
can be measured, in principle, in Primakov experiments or in
photoproduction experiments (apart from low-energy photon-photon
scattering). Such experiments have been performed but the present
situation is  somewhat confused,  e.g. the result  of MAMI~\cite{mami}
and the preliminary result from COMPASS~\cite{compass} are not in good
agreement, to mention only the most recent experiments.  

This paper  is motivated by  the new experimental measurements  by the
Belle    collaboration    of   $\gamma\gamma\to\pi\pi$    differential
cross-sections for both charged pions~\cite{bellemori2pic_1,bellemori2pic_2}
and   neutral  pions~\cite{belleuehara2pi0_1,belleuehara2pi0_2}.  Only
their  charged pion  results have  been used  in  previous theoretical
analysis. There has also been very significant progress,
recently,  in  measuring  the  $\pi\pi$ scattering  amplitude  at  low
energies by the NA48/2~\cite{NA48/2Kl4,NA48/2cusp} 
the DIRAC~\cite{DIRAC} and E865~\cite{E865} experiments.
We   will   focus  here   on   relating  the   $\gamma\gamma\to\pi\pi$
experimental results and the low-energy sector of QCD.  
We  will  argue  that  using chiral  constraints  is  useful  in
analyzing  the data  and that,  in return,  chiral information  can be
extracted from  the data. This  might appear puzzling at  first sight,
because  Belle's data  does  not  cover the  very  low energy  region:
the $\pi^0\pi^0$ data covers the range $E\gapprox 0.6$ GeV and the $\pi^+\pi^-$
data  the range $E\gapprox  0.8$ GeV.  Extrapolation is  possible
due  to  theoretical properties  of  scattering  amplitudes in  the
standard  model,  in  particular,   the  property  of  analyticity  of
partial-wave  amplitudes as a  function of  energy. This  property, we
recall,    is    a    proved    consequence    of    confinement    in
QCD~\cite{oehme}. Combining  with unitarity of  the $S$-matrix enables
one to  disentangle the effects  of the final-state interaction  by the
Muskhelishvili-Omn\`es  (MO)
method~\cite{omnes58,mushkebook}.  Application  to
$\gamma\gamma\to\pi\pi$  amplitudes was discussed for the first
time    in    ref.~\cite{gourdinmartin}.    Explicit    results    for
$\gamma\gamma\to  \pi^+\pi^-$  taking  into  account  current  algebra
constraints were obtained in~\cite{goblerosner1,goblerosner2}.
This was reconsidered a few years
later~\cite{morganpenning87,morganpenning88,donohol93} after the first
reliable experimental results in the low-energy region became
available. Refs.~\cite{morganpenning88,donohol93} also discuss how the
MO dispersive representation matches with the chiral one-loop
representation, which had been computed in
refs.~\cite{donohollin88,Bijnens:1987dc}  and eventually lead to a
parameter free prediction in the low-energy region. 

A simplifying feature of low energy is that $\pi\pi$ scattering can be
considered elastic. If one is interested in the 1 GeV region 
or slightly above, it becomes necessary to
take inelasticity into account. This is feasible due to a
specific feature of $\pi\pi$ scattering: the fact that inelastic
scattering to $4\pi$ or $6\pi$ states (which are not treatable by MO
methods)  are suppressed
in practice and can be neglected up to $E\simeq1.2-1.3$ GeV. The
remaining relevant inelastic channels, $K\bar{K}$ or $\eta\eta$, are
two-body channels which are perfectly treatable, in principle, by MO methods.  
In this paper, we take into account $\pi\pi\to K\bar{K}$ scattering,
which is particularly important in the $I=0$ $S$-wave near 1
GeV. This amplitude suffers from a long lasting unresolved
experimental discrepancy very near the  $K\bar{K}$ threshold which
limits the determination of the properties of the $f_0(980)$ scalar
meson. It is interesting that this discrepancy can be interpreted in
terms of two different scenarios for isospin breaking. We will discuss
the relevance of Belle's results near 1 GeV in eventually clarifying
this issue. 

Application of the multichannel MO method to $\gamma\gamma\to\pi\pi$
has been attempted first in ref.~\cite{sundermeyer} and then discussed
in some detail, on the basis of specific $T$-matrix parametrizations
for which the MO integral equations are analytically solvable, in
ref.~\cite{babelon76}. More recently, it was applied to Belle's data
on $\pi^+\pi^-$ in ref.~\cite{zheng09}. These authors have 
considered the extrapolation of the amplitude in the complex plane,
so as to define, and then extract, the couplings of the scalar mesons
$\sigma(600)$ and $f_0(980)$ to two photons. Determination of 
these couplings had aroused significant interest in the literature
(e.g.~\cite{mennessier83,boglionepennington,oller08,Mennessier:2008kk}, 
a more complete list can be found in~\cite{zheng09}). One motivation 
is to probe the structure of scalar mesons and identify the
glueballs.  There is some scatter in the results obtained. 
In the present paper, we concentrate on extrapolating on the
real axis, towards the low-energy region. This brings constraints
on the amplitude which should prove useful also for extrapolating away
from the real axis. This will be discussed elsewhere. 

The plan of the paper is as follows. After introducing notation
for the amplitudes and their partial-wave expansions we write the
unitarity equations in the one and two-channel approximations. 
Next, we formulate the MO-type dispersive
representations. Concerning the left-hand cut, we find it advantageous 
to treat differently the QED Born term and the multipion contributions
in the MO representations. The latter are kept in the form of a 
subtracted left-cut spectral integral. 
We then implement the (usual) approximation
of retaining only resonance contributions. All resonances with mass up
to $\approx 1.3$ GeV are included and we show that a certain regulation
operates between resonances of different spin and different
parity depending on the helicity states.  Subtractions at
$s=0$ are introduced in the MO dispersive representations
in order to suppress higher energy regions in the
integrands where our truncated unitarity equations no longer
apply. The subtraction constants have simple relations to dipole and
quadrupole pion polarizabilities and are to be determined from fits to
the data. Chiral constraints may be applied to the fit. This is
necessary because the errors in  Belle's data are completely dominated
by systematics and the usual statistical interpretation of the
$\chi^2$ does not apply, strictly speaking. Only one of the relevant
$p^6$ chiral coupling constants is known in a model independent way
from a chiral sum rule. We show that this information implies a
relation between dipole and quadrupole polarizabilities of the neutral
pion which we implement in the fit. 
Then, we describe our inputs for the $\pi\pi\to\pi\pi,K\bar{K}$
$T$-matrix  elements.  
Our subtracted dispersive integrals
emphasize the low-energy part of the integrands. We employ a parametrization
which allow for some freedom near the $K\bar{K}$ threshold since the
Belle data probes this region in some detail.
Finally, we  display 
comparisons between the fitted MO amplitudes and the experimental data
and discuss the implication for the pion polarizabilities and the $p^6$
chiral coupling constants. 

\section{Kinematics, unitarity relations}
\begin{figure}
\bc
\includegraphics[width=0.7\linewidth]{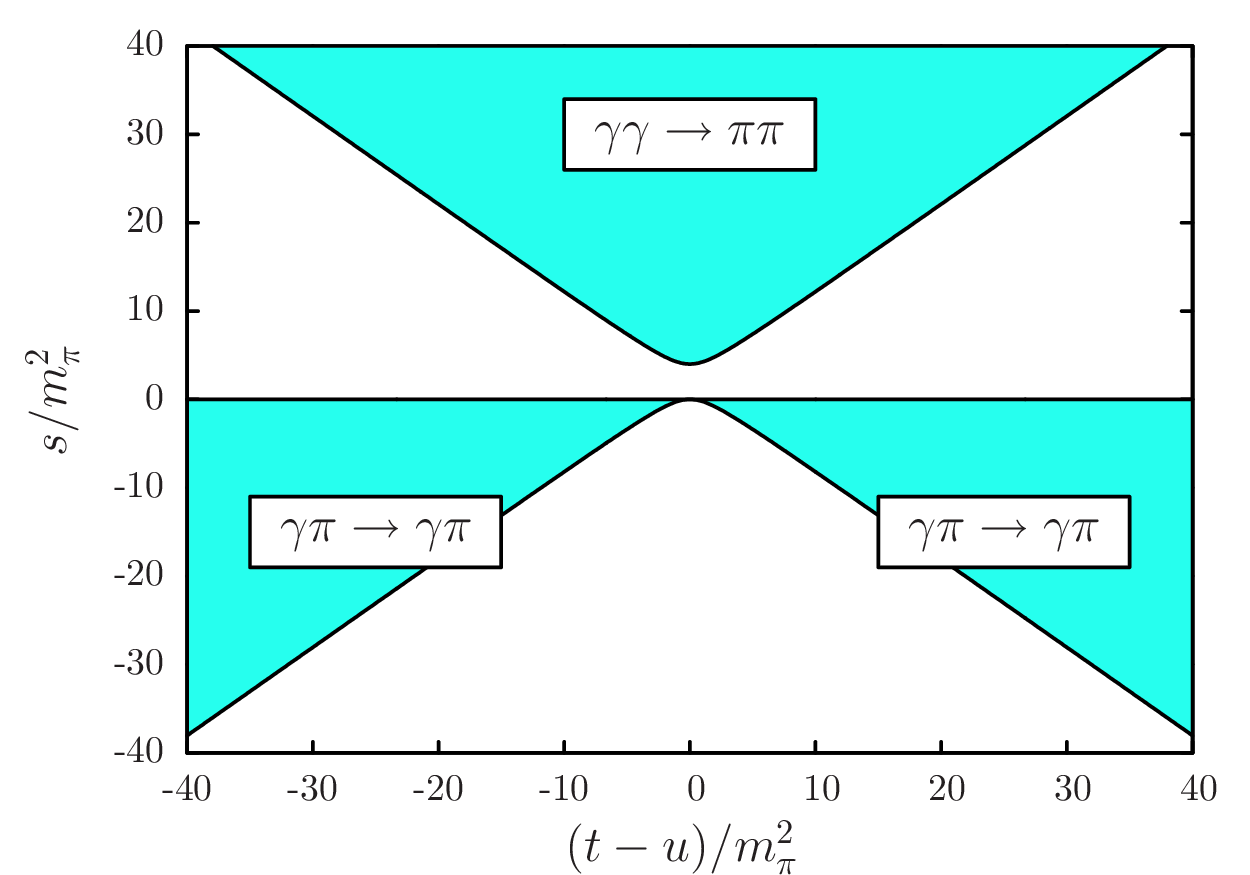}
\caption{\sl Mandelstam plane showing the physical regions for
$\gamma\gamma\to 2\pi$ and $\gamma\pi\to\gamma\pi$ processes.}
\label{fig:mandel2gams}
\ec
\end{figure}
We consider the processes
$\gamma(q_1,\lambda)\gamma(q_2,\lambda')\to \pi^+(p_1)\pi^-(p_2)$
or  $\pi^0(p_1)\pi^0(p_2)$  where  $\lambda(\lambda')=\pm 1$  are  the
photons       helicities.       We       will      also       consider
$\gamma(q_1,\lambda)\gamma(q_2,\lambda')\to K(p_1) \bar{K}(p_2)$ with
$I=0$ which plays a role in the $S$-wave via coupled channel
unitarity. We take the Mandelstam invariants as
\be
s=(q_1+q_2)^2,\quad t=(q_1-p_1)^2,\quad u=(q_1-p_2)^2
\en
which satisfy $s+t+u=2\mpid$.  The  physical regions in the Mandelstam
plane for $\gamma\gamma\to 2\pi$ and for the crossed-channel amplitude
$\gamma\pi\to\gamma\pi$ are shown in fig.~\fig{mandel2gams}.
The scattering angle $\theta$ in the two photon
center-of mass system is related as follows to $s$, $t$, $u$ 
\be
\cos\theta=\frac{t-u}{\sqrt{s(s-4\mpid)}}\ .
\en
We write the $S$-matrix element for $\gamma\gamma\to \pi^+\pi^-$ as
\be
\outleft \pi^+(p_1)\pi^-(p_2)\vert \gamma(q_1,\lambda)
\gamma(q_2,\lambda')\inright=i e^2(2\pi)^4\delta^4(P_f-P_i) 
\e^{i(\lambda-\lambda')\phi}\,H^c_{\lambda\lambda'}(s,t)
\en
factoring out $e^2$ as well as the explicit dependence on the 
azimuthal angle $\phi$.  With this convention, $H^c_{\lambda\lambda'}$ 
is a function of the Mandelstam variables. 
Similarly, in the case of $\gamma\gamma\to \pi^0\pi^0$ we denote the scattering
amplitude by $H^n_{\lambda\lambda'}$.
In the case of  $\gamma\gamma\to K\Kbar$ scattering
we denote the charged and neutral amplitudes 
by $K^c_{\lambda\lambda'}$ and $K^n_{\lambda\lambda'}$ respectively.
We assume that isospin is exactly conserved by the strong interaction.
It is then useful to consider the 
amplitudes which correspond to $\pi\pi$ or $K\Kbar$ final states with 
definite isospin $I$. We will label them as $H^I_{\lambda\lambda'}$ and
$K^I_{\lambda\lambda'}$. Because of parity conservation, 
only final states with even values of the angular momentum $J$
are allowed, $J=0,2,4,\cdots$ Invoking also charge conjugation invariance
the isospin values must be $I=0$ or $I=2$ in the case of $\pi\pi$, 
while in the case of $K\Kbar$ both $I=0$ and $I=1$
can couple to $\gamma\gamma$. The relations between the amplitudes 
$\gamma\gamma\to \pi^+\pi^-$, $\pi^0\pi^0$ and the isospin ones
$\gamma\gamma\to (\pi\pi)_{I=0,2}$ read
\be\lbl{Hisomatrix}
\left(\ba{r}
\sqrt2 H^c_{\lambda\lambda'} \\
H^n_{\lambda\lambda'}\ea\right)=
\left(\ba{rr}
-\sqrt{\frac{2}{3}} & -\sqrt{\frac{1}{3}}\\
-\sqrt{\frac{1}{3}} &  \sqrt{\frac{2}{3}}
\ea\right)
\left(\ba{l}
H^0_{\lambda\lambda'}\\
H^2_{\lambda\lambda'}
\ea\right) \ .
\en
In the case of kaons, the analogous relations read, 
\be\lbl{Kisomatrix}
\left(\ba{l}
K^c_ {\lambda\lambda'}\\
K^n_ {\lambda\lambda'} 
\ea\right)=
\left(\ba{rr}
-\sqrt{\frac{1}{2}} & -\sqrt{\frac{1}{2}}\\
-\sqrt{\frac{1}{2}} &  \sqrt{\frac{1}{2}}
\ea\right)
\left(\ba{l}
K^0_{\lambda\lambda'}\\
K^1_{\lambda\lambda'}
\ea\right)\ .
\en
It is  useful to carry out a tensorial decomposition of the 
photon-photon scattering amplitudes. Writing
\be
H_{\lambda\lambda'}(q_i,p_i)=\epsilon_1^\mu(\lambda) \epsilon_2^\nu(\lambda')\, 
W_{\mu\nu}(p_i,q_i)
\en
where $\epsilon_i$ are the polarization vectors of the photons. 
$W_{\mu\nu}$ can be decomposed as 
\be
W_{\mu\nu}= A(s,t,u) \,T_{1\mu\nu}+ B(s,t,u) \,T_{2\mu\nu} 
\en 
where
\bea
&&  T_{1\mu\nu}= {1\over2}s\,g_{\mu\nu}-q_{1\nu}q_{2\mu}\nonumber \\
&&  T_{2\mu\nu}=2s\Delta_\mu\Delta_\nu-(t-u)^2g_{\mu\nu}
-2(t-u)(q_{1\nu}\Delta_\mu -q_{2\mu}\Delta_\nu)
\ena
where $\Delta=(p_1-p_2)$. 
In this manner, the Ward identities are satisfied as follows,
\be
q_1^\mu W_{\mu\nu}= q_2^\nu W_{\mu\nu}=0\ . 
\en
The functions $A$ and $B$ satisfy analyticity properties as a function of 
$s$, $t$, $u$ and they are symmetric under crossing $(t,u)\to (u,t)$ 
(because of Bose symmetry of the two photon system). 
One can  express the helicity  amplitudes in terms  of $A$ and  $B$ as
follows\footnote{The  polarization vectors  are  chosen in  accordance
  with the phase convention of Edmonds~\cite{edmonds} for spherical tensors.},
\bea\lbl{tens2hel}
&&H_{++}=H_{--}=  {1\over2}s\,A -s(s-4\mpid)\,B\nonumber \\
&&H_{+-}= H_{-+}= 4(tu-\mpiq)\,B \ .
\ena
Finally, the  differential cross section  for $\gamma\gamma\to \pi\pi$
has the following expression
\be
{d\sigma\over d\Omega}= {\alpha^2\over8s}\,\beta_\pi(s) \left(
\vert H_{++}\vert^2 + \vert H_{+-}\vert^2 \right)
\en
with
\be
\beta_\pi(s)=\sqrt{1-{4\mpid\over s}}\ .
\en

\subsection{Partial-wave expansions}
In order to perform the partial-wave expansion for helicity amplitudes
we use the Jacob and Wick~\cite{jacobwick58} formulas,
\be
\braque{\theta\phi \lambda_c \lambda_d\vert T \vert 0,0,\lambda_a\lambda_b}
=N \sum_J (2J+1) D^{*J}_{\lambda_a-\lambda_b,\lambda_c-\lambda_d}(\theta,\phi)
\braque{\lambda_c \lambda_d\vert T_J\vert \lambda_a\lambda_b}
\en
where $N$ is a normalization factor which can be chosen arbitrarily.
Let us list below the partial-wave expansions for all the scattering amplitudes
which are relevant in our work,
\be\lbl{pwexpands}
\setlength{\arraycolsep}{1pt}
\begin{array}{rrrl}
\gamma\gamma\to \pi\pi:\phantom{K}
                   & H^I_{\lambda\lambda'}=& \sum (2J+1)& \,
h^I_{J,\lambda\lambda'}(s) \,d^J_{\lambda-\lambda',0}(\theta) \\[0.2cm]
\gamma\gamma\to K\Kbar: \phantom{K}
                   & K^I_{\lambda\lambda'}=& {1\over\sqrt2}\sum (2J+1)& \,
k^I_{J,\lambda\lambda'}(s) \,d^J_{\lambda-\lambda',0}(\theta) \\[0.2cm]
\pi\pi \to \pi\pi : \phantom{K}
                   & F^I= & 32\pi \sum (2J+1) & \, f_J^I(s)
\,d^J_{00}(\theta)\\[0.2cm]
\pi\pi \to K\Kbar: \phantom{K}
                   & G^I= & 16\sqrt2\pi\sum (2J+1)& \,
g^I_J(s) \,d^J_{00}(\theta)\\[0.2cm]
K\Kbar\to K\Kbar: \phantom{K}
                   &  R^0=   &  16\pi   \sum  (2J+1)  &   \,  r^0_J(s)
\,d^J_{00}(\theta)\ .
\end{array}
\en
The different normalization factors are chosen such as to ensure
simple formulas for the unitarity relations satisfied by the partial-wave
amplitudes.  
The general unitarity relation reads, for a given $T$-matrix element
\be\lbl{Tunitar}
T_{fi}-T^*_{if} =i\sum_n T^*_{nf} T_{ni}\ .
\en
If the  energy is sufficiently  small, the sum  over intermediate
states is limited to just  one state, $\pi\pi$ (elastic unitarity). We
will  consider this  to be  a  reasonably good  approximation for  our
purposes except in the $I=0$ $J=0$ case (see sec.~\sect{omrepres}).  At
the level of the partial-waves, if $J\ne0$ or $I\ne0$ we then have
\be\lbl{pwunitar}
\im h^I_{J,\lambda\lambda'}(s)= \theta(s-4\mpid) \beta_\pi(s) f^I_J(s)
h^{*I}_{J,\lambda\lambda'}(s)\ .
\en
For $I=0$ $J=0$ we also include $K\bar{K}$ in the sum~\rf{Tunitar}.
The unitarity relation can be written in matrix form
\be\lbl{pwunitar00}
\left(\ba{c}
\im h^0_{0,++}(s)\\
\im k^0_{0,++}(s)
\ea\right)= T\Sigma
\left(\ba{c}
h^{*0}_{0,++}(s)\\
k^{*0}_{0,++}(s)
\ea\right) \nonumber
\en
with
\be
T=\left(\ba{ll}
f^0_0(s) & g^0_0(s)\\ 
g^0_0(s) & r^0_0(s)
\ea\right)
\quad
\Sigma=\left(\ba{cc}
\theta(s-4\mpid) \beta_\pi(s) & 0 \\ 
0 & \theta(s-4\mkd) \beta_K(s)
\ea\right)\ .
\en

\section{(Modified) Omn\`es -Muskhelishvili representations}\lblsec{omrepres}
The partial-wave photon-photon scattering amplitudes 
$h^I_{J,\lambda\lambda'}(s)$ are analytic functions of the variable $s$ 
with two cuts on the real axis: 1) a right-hand cut extending from 
$4\mpid$ to $\infty$ and 2) a left-hand cut extending from $-\infty$
to $0$. The discontinuity along the right-hand cut is given 
by the unitarity relations \rf{pwunitar},~\rf{pwunitar00}. 
These properties are the basis of the 
Muskhelishvili-Omn\`es  (MO) method  for treating the final-state 
interaction problem, which has been applied
for the first time to the $\gamma\gamma\to \pi\pi$ scattering
amplitudes by Gourdin and Martin~\cite{gourdinmartin}. 
The usual method~\cite{omnes58} is based on writing a dispersion relation for 
the function,
\be\lbl{omstandard}
\tilde{F}(s)\equiv \Omega^{-1}(s) [ F(s) -F_L(s) ]
\en
where $F$  is the  amplitude of  interest, $F_L$  the part  of this
amplitude which has a left-hand cut and $\Omega$ is the Omn\`es
function.  The  function  $\tilde{F}(s)$,  by construction,  has  only  a
right-hand cut. 
We will use a slightly modified version here, which treats
on a different  footing the part of the  left-hand cut associated with
the   QED  Born   term  (which   is  known   exactly)  and   the  
remaining part, which we will associate with resonance exchanges.
Instead of ~\rf{omstandard}, we will consider
\be\lbl{ommodif}
\tilde{F}_{mod}(s)\equiv\Omega^{-1}(s) [ F(s) -F^{Born}_L(s) ]
\en
i.e. we subtract only the Born term piece such that the function 
$\tilde{F}_{mod}(s)$ has both a right-hand  cut and a left-hand cut. A
simplification  arising in  this  modified approach  is  that all  
polynomial terms  are absorbed into  the subtraction constants  of the
dispersive representation.  

In practice, in order to evaluate the imaginary part of 
$\tilde{F}(s')$ (or $\tilde{F}_{mod}(s')$) on the right-hand cut,
which is needed in the dispersion relation,
one   sets   $\im   [\Omega^{-1}(s')   F(s')]$  equal   to   zero   in
eqs.~\rf{omstandard}~\rf{ommodif}.  
This is  exact in  the energy region  where scattering is  elastic but
becomes inaccurate at higher energies.
%
The  influence of  this
inaccuracy can  be reduced by writing  down over-subtracted dispersion
relations,  so  as  to   suppress  the  integrand  in  the  inelastic
region. With this  in mind, it also makes sense  to define the Omn\`es
function  over  an infinite  range,  making a
plausible guess\footnote{To be more specific, 
we assume that $I=2$ phase-shifts tend to to zero. For $I=0$ and $J=0$
we take: 
$\lim_{s\to\infty}\delta_{\pi\pi}(s)=2\pi$, $\lim_{s\to\infty}\delta_{KK}(s)=0$ 
and $\lim_{s\to\infty}\vert T_{\pi\pi\to K\bar{K}}(s)\vert=0$ while for $I=0$ and
$J=2$ we assumed that the phase-shift goes to $\pi$.} 
concerning the behaviour of the phase-shift as $s'\to\infty$. 
Alternatively, one could perform  all the dispersive integrations over
a finite range $s'\le s_c$ and approximate the contributions 
from   the    range   $s'>   s_c$   by    polynomials   with   unknown
coefficients. This should be  practically equivalent to the procedure
adopted here. 
In the end, the sensitivity of the results on the higher 
energy ranges of the various integrals would have to be included in 
the errors. 

Let us now consider the MO representations in more detail. 

\vskip2mm
\noindent{\bf $\bullet$ $S$-wave $I=0$}\\
For  the  $I=0$  $S$-wave,  it  is necessary  to  generalize  the  MO
representation  to two  channels  in order  to  properly describe  the
$f_0(980)$  resonance  energy  region.  The Omn\`es  function  must  be
replaced by a $2\times 2$ Omn\`es matrix
\be\lbl{bbomega}
\overline{\overline{\Omega}}(s)=\left(
\ba{cc}
\Omega_{11}(s)\ \Omega_{12}(s)\\
\Omega_{21}(s)\ \Omega_{22}(s)
\ea\right)\ .
\en
The matrix elements of the Omn\`es matrix are analytic functions of $s$
with  only  a   right-hand  cut  as  in  the   one-channel  case.  The
discontinuities along this  cut are given, in terms  of the $2\times2$
$T$-matrix, by equations analogous to~\rf{pwunitar00} that read,
\be\lbl{om22disc}
\im\overline{\overline{\Omega}}(s)=T\Sigma 
\overline{\overline{\Omega}}^{\,*}(s)\ .
\en
Unlike   the   one-channel   case, the  MO  equations  have no known analytic
solutions for two or more channels~\cite{mushkebook}, but accurate numerical
solutions can be constructed~\cite{dgl,moussNf}. 
The MO representation couples 
$\gamma\gamma\to (\pi\pi)_{I=0}$ $S$-wave amplitude and the 
$\gamma\gamma\to  (K\bar{K})_{I=0}$ $S$-wave  amplitude.  We write,  a
priori, a representation which involves four subtraction parameters
\bea\lbl{2chanrepres}
&& \left(
\ba{l} 
h^0_{0,++}(s)\\
k^0_{0,++}(s)
\ea\right) =
\left(
\ba{l} 
\bar{h}^{0,Born}_{0,++}(s)\\
\bar{k}^{0,Born}_{0,++}(s)
\ea\right) + 
\overline{\overline{\Omega}}(s)\times\Bigg[
\left(\ba{l} 
b^{(0)} s +b^{'(0)} s^2\\
b_K^{(0)} s +b_K^{'(0)} s^2
\ea\right) 
\nonumber\\
&& \qquad
+{s^3\over\pi}\int_{-\infty}^{-s_0}{ds'\over(s')^3(s'-s)} 
\overline{\overline{\Omega}}^{-1}(s')\im
\left(\begin{array}{l} 
\bar{h}^{0,Res}_{0,++}(s')\\
\bar{k}^{0,Res}_{0,++}(s')
\end{array}\right)
\nonumber\\
&& \qquad
-{s^3\over\pi}
\int_{4\mpid}^\infty 
{ds'\over (s')^3(s'-s)} \im \overline{\overline{\Omega}}^{-1}(s')
\left(\begin{array}{l} 
\bar{h}^{0,Born}_{0,++}(s')\\
\bar{k}^{0,Born}_{0,++}(s')
\end{array}\right)
\Bigg]\ .
\ena
As usual, subtraction  constants have been set equal  to zero in order
to comply with the soft-photon theorem~\cite{low,abarbanel} near $s=0$ e.g. 
\be\lbl{sofphoton}
h^0_{0,++}(s)-\bar{h}^{0,Born}_{0,++}(s)=O(s)
\en
The left-cut functions $\bar{h}^{0,Born}_{0,++}(s')$, 
$\bar{k}^{0,Born}_{0,++}(s')$, $\bar{h}^{0,Res}_{0,++}(s')$, 
$\bar{k}^{0,Res}_{0,++}(s')$ which enter into this representation will
be given explicitly in sec.~\sect{leftcut}. 

Due to the property of real analyticity the discontinuity of the
amplitude $h^0_{0,++}(s)$  across the right-hand is  expressed in terms
of a phase $\phi_{00}$ 
\be\lbl{phi00def}
h^0_{0,++}(s+i\epsilon)=e^{2i\phi_{00}(s)} h^0_{0,++}(s-i\epsilon)
\qquad (s\ge 4\mpid)\ .
\en
This phase is  equal (modulo $\pi$) to the  $\pi\pi$ phase-shift below
the  $K\bar{K}$  threshold by  Watson's  theorem. The  representation
given above  provides a modelling  of $\phi_{00}$ above  the $K\bar{K}$
threshold  (depending on  the polynomial  parameters) which  is
plausible below the effective onset of $4\pi$ 
inelasticity.  The amplitude $h^0_{0,++}(s)$ satisfies a
one-channel Omn\`es  representation in  terms of the  Omn\`es function
associated  with $\phi_{00}$  and two  polynomial parameters.  We have
verified this property as a check of our numerical calculations.

\vskip2mm
\noindent{\bf $\bullet$ $S$-wave $I=2$}\\
In this case, $K\bar{K}$ inelasticity is not allowed. We
will then disregard inelasticity in the energy region of interest. We write an
MO representation with two subtraction constants
\bea\lbl{1chanrepres0}
&& h^2_{0,++}(s)= \bar{h}^{2,Born}_{0,++}(s) +
\Omega_0^2(s)\,\Bigg[ b^{(2)}s + b^{'(2)}s^2
\\
&& +{s^3\over\pi}\int_{-\infty}^{-s_0} {\im\bar{h}^{2,Res}_{0,++}(s')\over 
\Omega_0^2(s') (s')^3 (s'-s)} ds'
   +{s^3\over\pi}\int_{4\mpid}^\infty {\sin\delta^2_0(s')\, 
\bar{h}^{2,Born}_{0,++}(s')\over  \vert \Omega_0^2(s')\vert
(s')^3 (s'-s)} ds'\Bigg]\nonumber\ . 
\ena
Here $\delta_0^2(s)$  is the $I=2$  $\pi\pi$ phase-shift and  the Omn\`es
function is given in terms of $\delta_0^2(s)$ by
\be
\Omega_0^2(s)=\exp\left( {s\over\pi}\int_{4\mpid}^\infty {\delta_0^2(s')\over
s' (s'-s)}\,ds'\right)\ .
\en

\vskip2mm
\noindent{\bf $\bullet$ $D$-waves $I=0$ and $I=2$}\\
In the $I=0$ case, the partial-wave analysis performed 
by Hyams et al.~\cite{hyams73} found the inelasticity around the
$f_2(1270)$ peak  to be of the  order of 30\%.  The PDG~\cite{pdg} now
quotes a smaller  value, approximately 15\%, of which  only 5\% is due
to  $K\bar{K}$,  the remaining  part  being  due  to $4\pi$.  We  will
therefore not attempt  a coupled channel description in  this case and
essentially ignore the inelasticity. 
In the $I=2$ case, the  final-state interaction is very small and will
also ignore the inelasticity.  
The Omn\`es method differs from that discussed above 
only by the fact that we must account properly for centrifugal 
barrier factors.  At small energies, indeed, it is not difficult to see
that the amplitudes with $J=2$ should behave as follows,
\be
\ba{lr}
h^I_{2,++}(s)-\bar{h}^{I,Born}_{2,++}(s)\sim & s^2 (s-4\mpid)\ ,\\
h^I_{2,+-}(s)-\bar{h}^{I,Born}_{2,+-}(s)\sim & s (s-4\mpid)\ .\\
\ea
\en
This is implemented by multiplying the Omn\`es function by $(s-4\mpid)$ and
by setting the appropriate subtraction constants to zero in the dispersive 
representation. One then obtains,
\bea\lbl{1chanrepres2++}
&& h^I_{2,++}(s)= \bar{h}^{I,Born}_{2,++}(s) 
+\Omega^I_2(s)\,s^2(s-4\mpid)\times
\nonumber\\
&& \phantom{ h^I_{2,++}(s)=  } 
\Bigg[ c^{(I)} +
{s\over\pi}\int_{-\infty}^{-s_0} {\im\bar{h}^{I,Res}_{2,++}(s')\over 
\Omega^I_2(s') (s')^3 (s'-4\mpid) (s'-s)} ds'
\nonumber\\
&& \phantom{ h^I_{2,++}(s)= [c } 
+ {s\over\pi}\int_{4\mpid}^\infty {\sin\delta^I_2(s')\, 
\bar{h}^{I,Born}_{2,++}(s')\over  \vert \Omega^I_2(s')\vert
(s')^3 (s'-4\mpid)(s'-s)} ds'\Bigg]
\ena
and
\bea\lbl{1chanrepres2+-}
&& h^I_{2,+-}(s)= \bar{h}^{I,Born}_{2,+-}(s) 
+\Omega^I_2(s)\,s(s-4\mpid)\times
\nonumber\\
&& \phantom{ h^I_{2,+-}(s)=  } 
\Bigg[ d^{(I)} + 
{s\over\pi}\int_{-\infty}^{-s_0} {\im\bar{h}^{I,Res}_{2,+-}(s')\over 
\Omega^I_2(s') (s')^2 (s'-4\mpid) (s'-s)} ds'
\nonumber\\
&& \phantom{ h^I_{2,+-}(s)= [d } 
+ {s\over\pi}\int_{4\mpid}^\infty {\sin\delta^I_2(s')\, 
\bar{h}^{I,Born}_{2,+-}(s')\over  \vert \Omega^I_2(s')\vert
(s')^2 (s'-4\mpid)(s'-s)} ds'\Bigg]\ .
\ena
These representations involve four subtraction constants $c^{(I)}$ and 
$d^{(I)}$. 

Let us finally remark that polynomials can be introduced in MO
representations  in different  ways (for  instance  multiplying the
Omn\`es functions).  It is easy  to show that any  representation can
always  be recast  in the  form  given above.   These are  convenient
because the relations between the polynomial coefficients and the pion
polarizabilities are now particularly simple. 

\subsection{Sum-rules}\lblsec{sumrules}
Above, we  have written over-subtracted dispersion  relations in order
to suppress the contributions from  large values of $\vert s'\vert$ in
the integrands.  Such representations  can be valid  only in  a finite
energy domain since, in general, they lead to diverging amplitudes\footnote{
The asymptotic conditions implemented in the $T$ matrices imply
that  the $I=0$  and $I=2$  Omn\`es functions  behave  respectively as
$s^{-1}$ and $s^0$ at large $s$.}  
when $s\to\infty$.  A priori,  the exact asymptotic  behaviour of
the  partial-wave amplitudes  is  not known,  but $S$-matrix  
unitarity provides the following bound 
\be
\vert h^I_{J,\lambda\lambda'}(s)\vert \le
{16\pi\over\sqrt{\beta_\pi(s)}}\ 
\en
which implies that one could write  representations  with  fewer
subtraction constants if one assumes that the integrands are known
sufficiently  well. This amounts to expressing some of the
subtraction constants as sum rules.  Such sum rules are very simply
obtained from the representations  written above by requiring that the
most singular terms as $s\to\infty$ are cancelled. 
For example, the $D$-wave constants $c^{(I)}$, $d^{(I)}$  get expressed as
\be\lbl{cIdIsr}
 \left. c^{(I)}\right\vert_{SR}=L^{(I)}_{3,++} +R^{(I)}_{3,++},\quad
 \left. d^{(I)}\right\vert_{SR}=L^{(I)}_{2,+-} +R^{(I)}_{2,+-}
\en
with
\bea\lbl{srintegrals}
&& L^{(I)}_{n,\lambda\lambda'}={1\over\pi}
\int_{-\infty}^{-s_0} {\im\bar{h}^{I,Res}_{2,\lambda\lambda'}(s')\over 
\Omega^I_2(s') (s')^n (s'-4\mpid) } ds' 
\nonumber\\
&& R^{(I)}_{n,\lambda\lambda'}={1\over\pi}
\int_{4\mpid}^\infty {\sin\delta^I_2(s')\, 
\bar{h}^{I,Born}_{2,\lambda\lambda'}(s')\over  \vert \Omega^I_2(s')\vert
(s')^n (s'-4\mpid)} ds'\ .
\ena
A  useful  test  (which  we  will  perform) of  the  validity  of  the
integrands, in particular of the modelling of the left-hand cut, is to
verify that the values of the subtraction constants obtained from such
sum rules are  not significantly different from those  obtained from
fitting to the experimental data.  

\section{Left-hand cut }\lblsec{leftcut}
In order to proceed with the previous formulas we need to specify 
the left-hand cut pieces  of the $\gamma\gamma\to\pi\pi$ amplitudes (as well
as the analogous contributions to $\gamma\gamma\to K\bar{K}$ in the case $I=0$,
$J=0$). 
Quite   generally,  the   left-hand   cut  can   be  associated   with
singularities in the crossed-channel (i.e. 
$\gamma\pi\to\gamma\pi$) partial-waves. A derivation based on the Mandelstam
double-spectral      representation      can      be     found      in
ref.~\cite{gourdinmartin}. The first  cross-channel singularity is the
pion pole, followed by the unitarity cuts due to $2\pi$,
$3\pi$ etc... Here,  the pion pole contribution will  be exactly taken
into  account. The discontinuities  $\rho_{n\pi}(s)$ associated
with the  unitarity cuts  are calculable from  ChPT at small  $s$, but
they are strongly suppressed  in this region: $\rho_{3\pi}$ has chiral
order $p^6$ while $\rho_{2\pi}$ is even more suppressed and has chiral
order $p^8$.  Here, due to the lack  of detailed experimental  information on
the $\gamma\pi\to\gamma\pi$ partial-waves, we will content ourselves with simple
resonance approximations. 

\subsection{QED Born amplitudes }
Let us first consider the QED Born term contribution 
to $\gamma\gamma\to \pi^+\pi^-$ (and $\gamma\gamma\to K^+ K^-$).  
The standard result for the helicity amplitudes reads
\bea\lbl{BornHll}
&&H^{c,Born}_{++}= {2s\mpid\over (t-\mpid)(u-\mpid)}\quad
H^{c,Born}_{+-}= {2(tu-\mpiq)\over (t-\mpid)(u-\mpid)}\ .
\ena
The corresponding Born amplitudes for $\gamma\gamma\to K^+ K^-$ 
are, of course, the same, with $m_\pi$ replaced by $m_K$.
Performing the partial-wave projection according to eqs.~\rf{pwexpands}, 
one obtains for $J=0$ and $J=2$,
\bea\lbl{leftborn}
&& \bar{h}^{c,Born}_{0,++}(s)= {4\mpid\over s}\,L_\pi(s)\ ,\nonumber\\
&& \bar{h}^{c,Born}_{2,++}(s)= -{2\mpid\over s}
\left[\left(1-{3\over\beta^2_\pi(s)}\right)
L_\pi(s)+{6\over\beta_\pi^2(s)}\right]\ ,
\nonumber\\
&& \bar{h}^{c,Born}_{2,+-}(s)= {\sqrt{6}\over4}
\left[ \left( 1-{1\over\beta^2_\pi(s)}\right)^2 \beta^2_\pi(s) L_\pi(s)
-{2\over\beta^2_\pi(s)} +{10\over3} \right]
\ena
with
\be 
L_\pi(s)={1\over \beta_\pi(s)}\log{ 1+\beta_\pi(s)\over 1-\beta_\pi(s)}\ .
\en
We will also need the $J=0$ partial-wave amplitude for 
$\gamma\gamma\to K^+ K^-$ which, as a result of the
normalization~\rf{pwexpands} reads 
\be
\bar{k}^{c,Born}_{0,++}(s)= {4 \sqrt2 \mkd\over s}\,L_K(s)\ ,
\en
and the isospin $I=0$ and $I=2$ projections which are easily deduced
from~\rf{Hisomatrix},~\rf{Kisomatrix}. 

\subsection{Resonance contributions}
Let us review below the modelling of the left-hand cut as a sum
of resonance pole contributions. 
Resonances which can contribute must have
spin larger than or equal to 1. We will  consider vector, axial-vector,
tensor and axial-tensor contributions. We start by determining the form
of the amplitudes and the relation of the coupling constants to the radiative
decay widths. We will then discuss the phenomenological determination of
the coupling constants.

\vskip2mm
{\bf $\bullet$ Vector resonances}\\
We can start 
with a Lagrangian coupling a vector meson $V$ a pseudo-scalar meson 
$P$ (a pion or a Kaon) and a photon field of the form,
\be\lbl{lagVPgam}
{\cal L}_{VP\gamma}= eC_V \epsilon^{\mu\nu\alpha\beta} F_{\mu\nu} 
\partial_\alpha P V_\beta \ .
\en
After a small calculation we reproduce the result first 
obtained by Ko~\cite{Ko},
\be\lbl{leftV0}
W_V^{\mu\nu}={\tilde C_V\over m^2_V-t}
\left[ (s-4m_P^2-4t) T_1^{\mu\nu} +{1\over2} T_2^{\mu\nu}
\right] +(t\leftrightarrow u)
,\quad \tilde{C}_V ={1\over2} C^2_V \ .
\en
The relation between the coupling $\tilde{C}_V$ and the decay
width of the resonance reads,
\be
\Gamma_{V\to P \gamma} =\alpha \tilde{C}_V { (m_V^2-m_P^2)^3 \over 3 m_V^3}\ .
\en
Performing the partial-wave projections of these amplitudes
one finds for $J=0$ and $J=2$
\bea\lbl{leftparwV}
&&\bar{h}^{V}_{0,++}(s)= 4\tilde{C}_V\left[ 
- {m_V^2\over\beta_\pi(s)} L_V(s) + s 
\right]
\nonumber \\
&& \bar{h}^{V}_{2,++}(s)=\tilde{C}_V {2m_V^2\over\beta_\pi(s)}\left[ 
(1-3X_V^2(s))L_V(s) +6X_V(s)
\right]
\\
&& \bar{h}^{V}_{2,+-}(s)=\tilde{C}_V {\sqrt6\over4} s\beta_\pi(s)\left[
(1-X_V^2(s))^2\,L_V(s)
+{2\over3}X_V(s)(5-3X_V^2(s))
\right]\nonumber
\ena
where the logarithmic function $L_V$  reads
\be
L_V(s)=\log{X_V(s)+1\over X_V(s)-1},\quad 
X_V(s)= {2m_V^2-2\mpid +s\over s\beta_\pi(s)}.
\en
One remarks here that a term linear in $s$ appears in the expression
for the amplitude $\bar{h}^V_{0,++}$. This is an illustration of possible
polynomial ambiguities. Indeed, if one uses an antisymmetric tensor 
description of a vector particle (e.g.~\cite{egpr}) this term would appear
with the opposite sign. 
In the modified MO representation
we need only the imaginary parts of the partial-wave amplitudes
along the cut, which are free of any ambiguity.
The left-hand cut is contained in the function $L_V(s)$. 
Rewriting $X_V$ as
\be\lbl{xvsv}
X_V^2(s)=1+{ 4m_V^2 (s+s_V)\over s(s-4\mpid)}\ ,\quad 
s_V= {(m_V^2-\mpid)^2\over m_V^2}
\en
one deduces that the cut extends from $-\infty$ to $-s_V$.
One then derives the imaginary parts of the amplitudes
\bea\lbl{imleftparV}
&& {1\over\pi}\im\bar{h}^V_{0,++}(s)=-4\tilde{C}_V {m^2_V\over\beta_\pi(s)}
\theta(-s-s_V)\nonumber\\
&& {1\over\pi}\im\bar{h}^V_{2,++}(s)=-2\tilde{C}_V {m^2_V\over\beta_\pi(s)}
(3X^2_V(s)-1)\theta(-s-s_V)\nonumber\\
&& {1\over\pi}\im\bar{h}^V_{2,+-}(s)={\sqrt6\over4}
\tilde{C}_V\,s \beta_\pi(s)(X^2_V(s)-1)^2
\theta(-s-s_V)\ .
\ena
We remark that the sharp cutoff $\theta(-s-s_V)$ which appears in the imaginary
parts is associated with the narrow width approximation. Introducing a
finite width smoothes the cutoff in the region $s=-s_V$.
The corresponding contributions to $\gamma\gamma\to K\bar{K}$ 
partial-wave amplitudes (i.e. $\bar{k}^{V}_{0,++}$, $\bar{k}^{V}_{2,++}$
$\bar{k}^{V}_{0,+-}$) are obtained
by replacing in eqs.~\rf{leftparwV}  $m_\pi$ by $m_K$ 
and $\tilde{C}_V$ by $\sqrt2\tilde{C}_V$ (in accordance with the normalizations
defined in eqs.~\rf{pwexpands}).

\vskip2mm
$\bullet${\bf Axial-vector resonances}\\
We will designate generically the C-odd axial-vectors as $B$ and
the C-even ones by $A$. 
The following Lagrangian describes $B\to P\gamma$ coupling,
\be
{\cal L}_{B\to P\gamma}= e C_B F_{\mu\nu}\partial^\mu B^\nu P\ ,
\en
from which we easily deduce the amplitude $\gamma\gamma\to PP$ amplitude
corresponding to $B$ exchange,
\be\lbl{axialampl}
{W}_B^{\mu\nu}= {\tilde C_B\over m^2_A-t}
\left[ (s-4m_P^2+4t) T_1^{\mu\nu} +{1\over2} T_2^{\mu\nu}
\right] +(t\leftrightarrow u)\ ,\quad \tilde{C}_B={1\over8}C^2_B\ .
\en
with
\be
\en
As before, the couplings $\tilde{C}_B$ 
must be deduced from experimental data
on $B\to P\gamma$ decay using
\be
\Gamma_{B\to P \gamma} =\alpha \tilde{C}_B { (m_B^2-m_P^2)^3 \over 3 m_B^3}\ .
\en
Working out the helicity amplitudes  
corresponding to eq.~\rf{axialampl}, one
sees that they are simply related to the helicity amplitudes associated
with vector resonances:
\be
\ba{ll}
H_{++}^B( \tilde{C}_B, m_B ,s,t)=  &-
H_{++}^V( \tilde{C}_V\to \tilde{C}_B\ , 
m_V\to m_B ,s,t) \\
H_{+-}^B( \tilde{C}_B, m_B ,s,t)=  &\phantom{-}
H_{+-}^V( \tilde{C}_V\to \tilde{C}_B\ , 
m_V\to m_B ,s,t) \\
\ea
\en
from which one easily deduces the imaginary parts 
of the partial-wave projections from~\rf{imleftparV}.

\vskip2mm
\noindent{\bf $\bullet$ Tensor resonances}\\
The Lagrangian coupling  a tensor meson (quantum numbers $2^{++}$)
to a photon and a pseudo-scalar 
meson must have the following form,
\be
{\cal L}_{TP\gamma}=e C_T\, \epsilon^{\mu\nu\alpha\beta} F_{\mu\nu} 
T_\alpha^{\phantom{\alpha}\lambda}\partial_\lambda\partial_\beta P\ .
\en
From this, we can first deduce the relation
between the decay width and the coupling constant
\be
\Gamma_{T\to P\gamma}= \alpha\,\tilde{C}_T { (m_T^2-m_P^2)^5\over
  5m_T^5}\ , \quad  \tilde{C}_T={C_T^2\over16}\ .
\en
After a small calculation we can obtain the form of the 
$\gamma\gamma\to PP $ amplitude generated by tensor meson
exchange
\be
W_T^{\mu\nu} = A_T(s,t) T_1^{\mu\nu} + B_T(s,t)T_2^{\mu\nu} 
+(t\leftrightarrow u)
\en
with
\bea
&& B_T(s,t)= {\tilde{C}_T \left[ (t+\mpid)^2+4m_T^2(s-\mpid)
\right]\over2 m_T^2 (m_T^2-t) }
\nonumber\\
&& A_T(s,t)= 2(s-4\mpid-4t)B_T(s,t) -{8\tilde{C}_T (t-\mpid)^2\over
  (m_T^2-t)}\ .
\ena
From here, we can  construct the helicity amplitudes and their 
projections. We quote below the imaginary parts of these 
along the left-hand cut
\bea\lbl{imleftparT}
&& {1\over\pi}\im \bar{h}^{T}_{0,++}(s)= 
-4\tilde{C}_T {m_T^2\over\beta_\pi(s)} 
\left[4s+3s_T)\right]\theta(-s-s_T)
 \\
&&{1\over\pi}\im \bar{h}^{T}_{2,++}(s)=-2\tilde{C}_T {m_T^2\over\beta_\pi(s)} 
(3X_T^2(s)-1) \left[ 4s+3s_T\right]\theta(-s-s_T)
\nonumber \\
&& {1\over\pi}\im \bar{h}^{T}_{2,+-}(s)={\sqrt6\over 4 }\tilde{C}_T\,
s\beta_\pi(s)  (X_T^2(s)-1)^2 \left[ 4s + s_T)\right] 
\theta(-s-s_T)\ .\nonumber
\ena

\vskip2mm
\noindent{\bf $\bullet$ Axial-tensor resonances}\\
For illustrative purposes, finally, let us consider axial-tensor
(i.e. with quantum number $J^P=2^{-}$) resonances.
The relevant Lagrangian has the following form
\be
{\cal L}_{T_AP\gamma}=e C_{T_A}\, F_{\mu\nu} T_A^{\nu\lambda} \partial_\mu
\partial_\lambda P
\en
and we deduce the following relation with the radiative decay width
\be
\Gamma_{T_A\to P\gamma}= \alpha\,\tilde{C}_{T_A} 
{ (m_{T_A}^2-m_P^2)^5\over 5m_{T_A}^5}\ ,
\quad  \tilde{C}_{T_A}={C_{T_A}^2\over64}\ .
\en
Next, one computes the diagrams contributing to $\gamma\gamma\to PP$
and one finds the result
\bea
&& B_{T_A}(s,t)= {\tilde{C}_{T_A} \left[ (t+\mpid)^2+4m_{T_A}^2(s-\mpid)
\right]\over2 m_{T_A}^2 (m_{T_A}^2-t) }
\nonumber\\
&& A_{T_A}(s,t)=2(s-4\mpid+4t) B_{T_A}(s,t) +
{8 \tilde{C}_{T_A}(t-\mpid)\over (m_{T_A}^2-t) }
\ena
Constructing the helicity amplitudes, one notices the following
simple relations between the tensor and the axial-tensor amplitudes 
\bea
&& H^{T_A}_{++}(\tilde{C}_{T_A},m_{T_A},s,t) = 
-H^{T}_{++}(\tilde{C}_{T}\to \tilde{C}_{T_A},m_{T}\to m_{T_A},s,t) \nonumber \\
&& H^{T_A}_{+-}(\tilde{C}_{T_A},m_{T_A},s,t) = 
H^{T}_{+-}(\tilde{C}_{T}\to \tilde{C}_{T_A},m_{T}\to m_{T_A},s,t)
\ena
Alternative expressions can be derived for the resonance exchanges, which
are somewhat more transparent physically and easier to generalize
to arbitrary spin $S$. They involve rotation functions
$d^S_{\lambda,\lambda'}(z_R)$  where $z_R $ is the center-of-mass
scattering angle  for the crossed-channel amplitude 
$\gamma\pi\to \gamma\pi$ at the resonance mass,
\be
z_R= \left. 1+{2st\over (t-\mpid)^2}\right\vert_{t=m_R^2} = 1+ {2s\over s_R}
\en
(using ~\rf{xvsv})
and generalized Legendre polynomials $P_J^m(X_R)$. The vector-exchange
partial-wave helicity amplitudes can be expressed as 
\bea
&& {1\over\pi}\im \bar{h}^{V}_{0,++}(s)= -4\tilde{C}_V  { m_V^2 s_V\over
\sqrt{s(s-4\mpid)} }P_0^0(X_V)\, d^1_{1,-1}(z_V)
\theta(-s-s_V)\nonumber \\
&& {1\over\pi}\im \bar{h}^{V}_{2,++}(s)= -4\tilde{C}_V  { m_V^2 s_V\over
\sqrt{s(s-4\mpid)} }P_2^0(X_V)\, d^1_{1,-1}(z_V)
\theta(-s-s_V)
\nonumber \\
&& {1\over\pi}\im \bar{h}^{V}_{2,+-}(s)= -{\sqrt6\over3} \tilde{C}_V 
{m_V^2 s_V \over\sqrt{s(s-4\mpid)} } P_2^2(X_V)\, d^1_{1,1}(z_V)
\theta(-s-s_V)\ .  
\ena
Analogously, the tensor exchange amplitudes involve the rotation functions
$d^2_{\lambda,\lambda'}(z_T)$
\bea
&& {1\over\pi}\im \bar{h}^{T}_{0,++}(s)= -4\tilde{C}_T 
{m^2_T s^2_T\over\sqrt{s(s-4\mpid)}} P_0^0(X_T)\,d^2_{1,-1}(z_T)
\theta(-s-s_T)
\nonumber \\
&&{1\over\pi}\im \bar{h}^{T}_{2,++}(s)= -4\tilde{C}_T 
{m^2_T s^2_T\over\sqrt{s(s-4\mpid)}} P_2^0(X_T)\,d^2_{1,-1}(z_T)
\theta(-s-s_T)\nonumber \\
&& {1\over\pi}\im \bar{h}^{T}_{2,+-}(s)= -{\sqrt6\over3} \tilde{C}_T 
{m^2_T s^2_T\over\sqrt{s(s-4\mpid)}} P_2^2(X_T)\,d^2_{1,1}(z_T)
\theta(-s-s_T)\ .
\ena
From these expressions, it is easy to guess the form of the amplitudes 
for an  arbitrary angular momentum $J$, or
generated by the exchange of a meson of arbitrary spin $S$. Because
of the polynomial functions $d^S_{\lambda,\lambda'}(1+2s/s_R)$, the behaviour
of the amplitudes as a function of $s$ becomes worse as the spin $S$ 
of the exchanged resonance increases. As usual, one expects that a
Regge-type regularization will occur upon including an infinite set of
resonances. In practice, one can simulate this by introducing a cutoff
on the left-cut integration.

\begin{figure}
\hskip-1cm\includegraphics[width=1.05\linewidth]{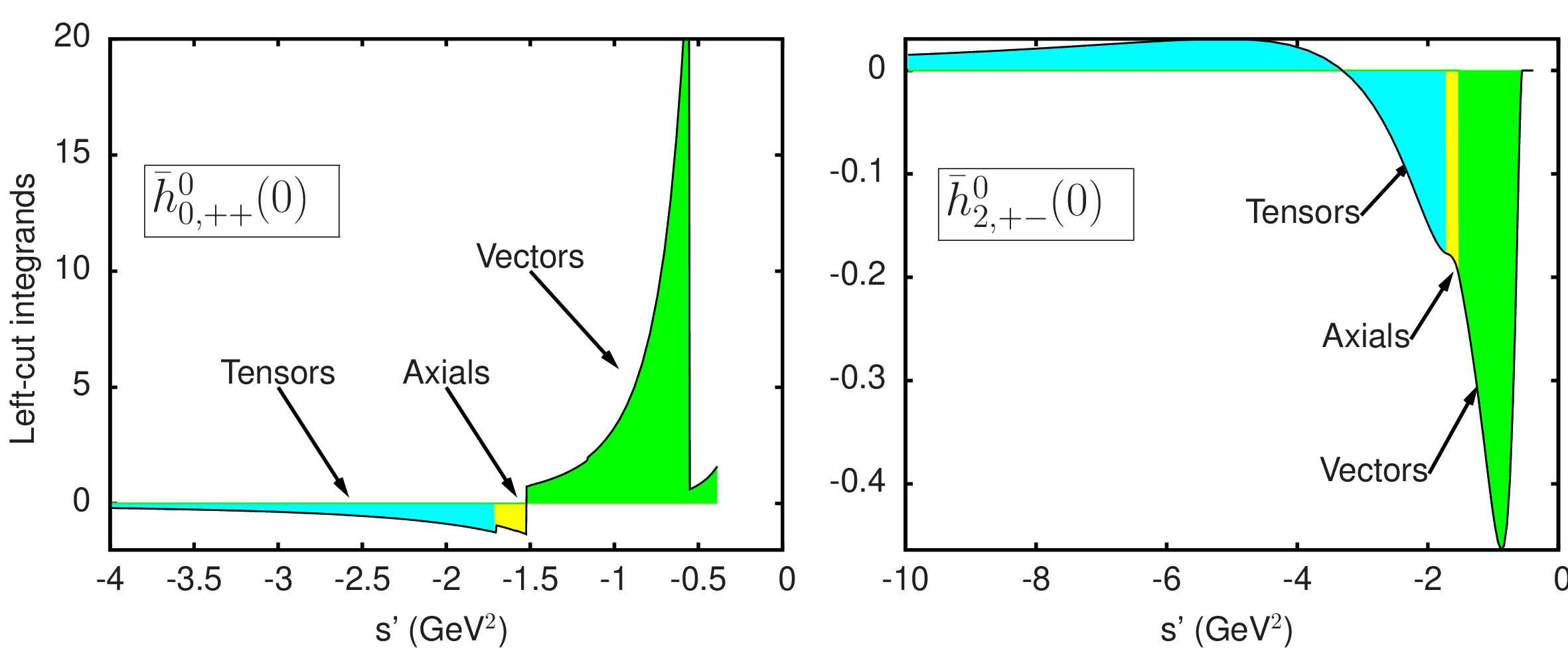}
\bc
\caption{\sl Left-cut integrand for $h^0_{0,++}(0)$ (left figure, see
  eq.~\rf{2chanrepres}) and  
$h^0_{2,+-}(0)$ (right figure, see eq.~\rf{1chanrepres2+-}) illustrating
the role of the various contributing resonances.}
\label{fig:hbarverif}
\ec
\end{figure}

\begin{table}[ht]
\bc
\bt{|c|llll|}\hline
$ \lambda\lambda'$ & V  & A   & T   & T$_A$ \\ \hline
$++$                &$-$ & $+$ & $+$ & $-$   \\
$+-$                &$-$ & $-$ & $+$ & $+$   \\ \hline  
\et
\caption{\sl Signs of the imaginary parts of the 
helicity amplitudes as generated from various
resonance exchanges}
\lbltab{signs}
\ec
\end{table}
Let us make a remark on the signs of these resonance
amplitudes. Varying $s$ from $-\infty$ to $-s_R$, the quantity
$X_R(s)$ varies from $-1$ to $1$. The Legendre polynomial $P_2(X_R)$ passes
through a zero, implying that $\im\bar{h}^R_{2,++}$ changes sign while
the amplitudes $\im\bar{h}^R_{2,+-}$ and  $\im\bar{h}^R_{0,++}$ do
not. This feature partly explains why, upon integration over $s$, the $+-$
$D$-wave amplitude is larger than the $++$ one. Furthermore,
there are alternating signs between the various resonance contributions: see 
table~\Table{signs}.  The table shows that for the $++$
amplitudes the signs alternate between resonances of different spin
and between $P$-even and $P$-odd resonances
while for helicity $+-$ amplitudes the signs alternate between
resonances of different spin. 
The resulting behaviour of some of the integrands from
which the left-cut functions are computed is illustrated 
in fig.~\fig{hbarverif}. This figure illustrates the numerical 
importance of the tensor contribution in the helicity $+-$ amplitude.

\subsection{Phenomenological determination of the coupling constants: }
\noindent{\bf $\bullet$ Neutral resonances}\\
The neutral resonances which can decay into a photon and a $\pi^0$ 
must be odd under charge conjugation. 
This is the case of the vector mesons
$\rho^0$, $\omega$ and their properties
are rather well known experimentally. The results from the PDG~\cite{pdg} and
the corresponding values of the couplings $\tilde{C}_V$ are collected in
table~\Table{neutralvec} below including also the result for the
$K^{*0}(892)$.
\begin{table}[hb]
\begin{center}
\bt{|l| ll|}\hline
\TT    &$\Gamma$ (KeV) & $\tilde{C}_R $ (GeV$^{-2}$)  \BB \\ \hline
$\omega\to\pi^0\gamma$ &  $703\pm25$   & $0.66\pm0.023$         \\
$\rho^0\to\pi^0\gamma$ &  $89\pm12$   &  $0.09\pm0.01$        \\
$\phi \to\pi^0\gamma$ &   $5.4\pm0.5$  & $(0.2\pm0.02)10^{-2}$         \\
$K^{*0} \to K^0\gamma$ &  $117\pm10$   & $0.20\pm0.02$         \\ \hline\hline
$h_1(1170)\to\pi^0\gamma$ &  $-$            & $\simeq0.45$         \\
$b_1(1235)\to\pi^0\gamma$ &   $-$          &  $\simeq0.05$        \\
$K_1(1270) \to K^0\gamma$ &    $73\pm 29$  & $0.024\pm0.010$         \\
$K_1(1400)\to K^0\gamma$ &    $280\pm 46$   & $0.063\pm0.010$         \\ \hline
\et
\caption{\sl Radiative widths of neutral
vector mesons and of  neutral $C-$odd axial-vector mesons
from the PDG~\cite{pdg} and the corresponding couplings $\tilde{C}_R$.}
\lbltab{neutralvec}
\end{center}
\end{table}

The experimental information concerning the C-odd axial meson radiative decays
is not as detailed as in the case of the vector mesons. The PDG quotes a
result for $b_1^+(1235)$ decay: $\Gamma(b_1^+(1235)\to \pi^+\gamma)=
240\pm60$ KeV, whereas the corresponding radiative widths of the neutral 
axials $b_1^0(1235)$ and $h_1(1170)$ have not yet been measured.   
A rough estimate of these, using nonet symmetry, is:
\be
\tilde{C}_{b_1^0(1235)}\simeq  \tilde{C}_{b_1^+(1235)},\quad
\tilde{C}_{h_1^0(1170)}\simeq 9\tilde{C}_{b_1^+(1235)}\ .
\en
Concerning the strange axials, we can use here the experimental results
on the radiative decays of the neutral $K_1(1270)$, $K_1(1400)$ which
can be found in the PDG.
We collect this information and the results for the $\tilde{C}_B$
couplings in the lower part of table~\Table{neutralvec}.
Finally, both $C$-odd and $C$-even axial-tensor mesons are expected to exist
in the quark model with a mass around 1.7 GeV~\cite{godfreyisgur}. 
Experimentally, the $C$-even axial-tensor meson $\pi_2(1670)$ is mentioned
in the PDG, but not the $C$-odd. The decay width of the $\pi_2(1670)$ into
$\gamma\pi$ is not known.

\vskip2mm
\noindent{\bf $\bullet$ Charged resonances}\\
Here, we can have contributions from charged vector mesons 
$\rho^+(770)$, $K^{*+}(892))$ and
from charged C-odd and C-even axial-vector mesons  
$a_1^+(1260)$, $ b_1^+(1235)$ ,$K_1^+(1270)$, $K_1^+(1400)$. 
We have also considered
the contributions from charged tensor mesons $a_2^+(1320)$, $K^{*+}_2(1430)$
since their masses are comparable to those of the axial-vectors. 
The relevant couplings $\tilde{C}_V$ for the vector mesons
in the charged case can be deduced from experiment and 
are collected in table~\Table{chargedvec}.
\begin{table}[h]
\bc
\bt{|l|ll|}\hline
\TT \     &$\Gamma$ (KeV) & $\tilde{C}_R$ (GeV$^{-2}$)   \\ \hline
$\rho^+\to\pi^+\gamma$ &  $68\pm7$   &  $0.066\pm 0.007$        \\
$K^{*+} \to K^+\gamma$ &  $50\pm5$   &  $0.085\pm 0.009$ \\ \hline\hline
$a_1^+(1260)\to\pi^+\gamma$ &  $640\pm240$   &  $0.15\pm0.06$        \\
$b_1^+(1235)\to\pi^+\gamma$ &  $230\pm60$    &  $0.05\pm0.01$        \\
$K_1^+(1270) \to K^+\gamma$ &  $-$   &  $\simeq 0.20$        \\ 
$K_1^+(1400) \to K^+\gamma$ &  $-$   &  $\simeq 0.00$        \\ \hline\hline
$a_2^+(1320)\to\pi^+\gamma$ & $287\pm30$ &   $0.052\pm0.005$        \\
$K^{*+}_2(1430)\to K^+\gamma$ & $241\pm50$ & $0.053\pm0.011$ \\ \hline
\et
\caption{\sl Same as table~\Table{neutralvec} for charged
vectors, axial-vectors and tensor resonances. }
\lbltab{chargedvec}
\ec
\end{table}
In the case of the axial-vectors, 
results are available for the $b_1^+(1235)$~\cite{b1exper} 
as well as for the $a_1^+(1260)$~\cite{a1exper} from Primakoff
experiments.
We must however keep in mind that results from 
photoproduction experiments~\cite{a1photo1,a1photo2} suggest
that the radiative width of the $a_1^+(1260)$ could actually be smaller than 
claimed in ~\cite{a1exper}.
Concerning the charged strange axials, $K_1(1270)$, $K_1(1400)$, 
unfortunately, no experimental information is available on their radiative 
widths. Rough estimates can again be made using nonet symmetry, 
which leads to the following relations
\bea
&&\tilde{C}_{K_1^0(1270)}+  \tilde{C}_{K_1^0(1400)}= 4 \tilde{C}_{b_1^+(1235)}
\nonumber\\
&&\tilde{C}_{K_1^+(1270)}+  \tilde{C}_{K_1^+(1400)}= \tilde{C}_{b_1^+(1235)}
+ \tilde{C}_{a_1^+(1260)}\ .
\ena
The first relation is obeyed by the experimental results within a factor of 
two. In order to determine the couplings $\tilde{C}_{K_1^+(1270)}$ 
and $\tilde{C}_{K_1^+(1400)}$ separately we note that one of them should be 
enhanced relative to the other  by the Lipkin mechanism~\cite{lipkin77}.  
It seems plausible that it should be $\tilde{C}_{K_1^+(1270)}$ because 
its main decay mode is via $K^+\rho^0$ which can produce
$K^+\gamma$ via vector meson dominance.

\section{Pion polarizabilities and chiral symmetry 
constraints}\lblsec{polarizab}
\subsection{Polarizabilities}
Polarizabilities are important observables which probe the 
structure of the pion. Using 
crossing symmetry and analyticity one can relate the amplitudes
$\gamma\gamma\to \pi^+\pi^-, \pi^0\pi^0$ at $t=\mpid$ and 
small $s$ to the electric and magnetic polarizabilities 
of the charged and the neutral pion respectively. 
These pionic observables have been computed in 
Chiral perturbation theory at next to leading
order~\cite{donohollin88,Bijnens:1987dc} and then at
next-to-next to 
leading order~\cite{bellucci,burgi,gasserivan05,gasserivan06}.
Some of these results which are relevant to our analysis will
be recalled in the next subsection. 
There are simple relations between the subtraction constants 
introduced in the dispersive relations for the partial-waves 
and the electric/magnetic polarizabilities of the pion.
From the partial waves with $J=0$ and $J=2$ which are involved 
in our analysis we can access the dipole and the quadrupole 
polarizabilities. We list below the relevant formulas. 
The polarizabilities of the neutral pion are defined as follows from the 
expansions  of  $H^n_{++}$  and  $H^n_{+-}$  around  $s=0$,  $t=\mpid$
(e.g.~\cite{gasserivan05}) 
\bea\lbl{polardefn}
&& {2\alpha\over\mpi} {H^n_{++}(s,t=\mpid)\over s}
=(\alpha_1-\beta_1)_{\pi^0} 
+{s\over12}(\alpha_2-\beta_2 )_{\pi^0} +\cdots\nonumber \\
&& {-2\alpha\over\mpi} {H^n_{+-}(s,t=\mpid)\over s}
=(\alpha_1+\beta_1)_{\pi^0} 
+{s\over12}(\alpha_2+\beta_2 )_{\pi^0}+\cdots
\ena
(the minus  sign in from of  $H^n_{+-}$ is associated  with our choice
for the photon polarization vectors). 
The charged pion polarizabilities are defined by the expansion of the 
amplitudes after removing the Born term, i.e. defining
$\hat{H}^c_{\lambda\lambda'}= {H}^c_{\lambda\lambda'}- 
{H}^{c,Born}_{\lambda\lambda'}$,
\bea\lbl{polardefp}
&& {2\alpha\over\mpi} 
{\hat{H}^c_{++}(s,t=\mpid)\over s}
=(\alpha_1-\beta_1)_{\pi^+} 
+{s\over12}(\alpha_2-\beta_2 )_{\pi^+} +\cdots\nonumber \\
&& {-2\alpha\over\mpi} 
{\hat{H}^c_{+-}(s,t=\mpid)\over s}
=(\alpha_1+\beta_1)_{\pi^+} 
+{s\over12}(\alpha_2+\beta_2 )_{\pi^+}+\cdots
\ena
Performing the partial-wave expansion, the $J=0$ and $J=2$ partial-waves
are  the only  ones  which  contribute to  the  dipole and  quadrupole
polarizabilities.

The polarizabilities  are simply related to  the subtraction constants
in the dispersive representations. Comparing with 
those~\rf{2chanrepres}~\rf{1chanrepres0} for $h^0_{0,++}$
and $h^2_{0,++}$ we obtain for the polarizability differences
\bea
&&(\alpha_1-\beta_1)_{\pi^+}=-{1\over\sqrt6}{2\alpha\over m_\pi}
(\sqrt2 b^{(0)} +b^{(2)})
\nonumber\\
&&(\alpha_1-\beta_1)_{\pi^0}=-{1\over\sqrt3}{2\alpha\over m_\pi}
(b^{(0)}-\sqrt2 b^{(2)}) \ .
\ena
Similarly, the polarizability sums are obtained from 
the subtraction constants $d^I$ 
which appear in the $J=2$ spin-flip amplitudes  $h^I_{2,+-}$ 
(see~\rf{1chanrepres2++},~\rf{1chanrepres2+-})
\bea
&&(\alpha_1+\beta_1)_{\pi^+}=-10\,\alpha m_\pi
(\sqrt2 d^{(0)} + d^{(2)} )
\nonumber\\
&&(\alpha_1+\beta_1)_{\pi^0}=-10\sqrt2\,\alpha m_\pi
(d^{(0)} -\sqrt2 d^{(2)} ) \ .
\ena
Considering the quadrupole polarizabilities now, we list below 
the polarizabilities defined for $I=0$ and $I=2$ amplitudes 
which can be easily combined using eq.~\rf{Hisomatrix}. For the
polarizability differences, we get
\bea
&& (\alpha_2-\beta_2)^{(0)}={24\alpha\over\mpi}\left(
\dot\Omega_{11}\,b^{(0)} + \dot\Omega_{12} b^{(0)}_K + b^{'(0)} 
+10\,\mpid c^{(0)}
\right)
\nonumber\\
&& (\alpha_2-\beta_2)^{(2)}={24\alpha\over\mpi}\left(
\dot\Omega^2_0\,b^{(2)} + b^{'(2)} +10\,\mpid c^{(2)}
\right)\ .
\ena 
For the sum of the quadrupole polarizabilities, we get
\bea\lbl{quadrupolesum}
&& (\alpha_2+\beta_2)^{(0)}=120\sqrt6\, 
\mpi\alpha \left( \dot\Omega^0_2\, d^{(0)}
+ d^{'(0)} \right)
\nonumber\\
&& (\alpha_2+\beta_2)^{(2)}=120\sqrt6\, 
\mpi\alpha \left( \dot\Omega^2_2\, d^{(2)}
+ d^{'(2)} \right)\ .
\ena 
In these formulas  the derivatives of the Omn\`es functions at $s=0$
appear, denoted by,  e.g. $\dot\Omega_{11}$. In eq.~\rf{quadrupolesum}
the quantities $d^{'(0)}$, $d^{'(2)}$ are given by sum rules
\be
d^{'(I)}=L^{(I)}_{3,+-}+R^{(I)}_{3,+-}
\en
where the integrals are defined in eq.~\rf{srintegrals}. 
Finally, the subtraction parameter $b^{(0)}_K$ which appears in the analysis
through coupled channel unitarity is related to the $I=0$ kaon polarizability 
as follows
\be
(\alpha_1 -\beta_1)_K^{(0)}= {\sqrt2 \alpha \over m_K} b^{(0)}_K\ .
\en

\subsection{Constraints from chiral symmetry}\lblsec{constraintch}
Chiral symmetry constrains the amplitudes 
$\gamma\gamma\to\pi^0\pi^0, \pi^+\pi^-$ for small values of the Mandelstam 
variables $s$, $t$. Computations up to NNLO in the chiral expansion
have been performed~\cite{bellucci,burgi,gasserivan05,gasserivan06}. At this
order, the amplitudes involve 13 coupling constants from the $O(p^6)$ chiral
Lagrangian. More precisely, three combinations of such couplings are involved
which we will denote $a_1^r$, $a_2^r$, $b^r$ for $\pi^0\pi^0$ and
$\tilde a_1^r$, $\tilde a_2^r$, $\tilde b^r$ for $\pi^+\pi^-$. Most of the 
$O(p^6)$ couplings are as yet undetermined, but in the case of $\pi^0\pi^0$ 
we can make use of a chiral sum rule for one coupling. We
explain this below. We also recall the chiral expressions for the pion 
polarizabilities which allows one to assess which chiral constraints
can be used in the fits to the experimental data.

\vskip0.2cm
\noindent{\bf a) $\pi^0\pi^0$:}\\
The dipole polarizabilities of the neutral pion have the following 
expressions in ChPT, in terms of the coupling combinations 
$a_1^r$, $a_2^r$, $b^r$
\bea
&& (\alpha_1-\beta_1)_{\pi_0}= {\alpha\over 16\pi^2 \fpid \mpi}\left[
-{1\over3} +{\mpid\over16\pi^2 \fpid}\left( a_1^r +8 b^r + X_{1-}(\mu)\right)
\right]\nonumber\\
&&  (\alpha_1+\beta_1)_{\pi_0}={\alpha\mpi\over (16\pi^2 \fpid)^2 }\left[ 
8 b^r  + X_{1+}(\mu) \right]
\ena
where the quantities $X_{1-}(\mu)$, $X_{1+}(\mu)$ involve chiral logarithms 
and known $O(p^4)$ couplings. Their detailed expressions can be found in
ref.~\cite{gasserivan05}. The chiral expressions for the quadrupole
polarizabilities read
\bea
&& (\alpha_2-\beta_2)_{\pi_0}= {\alpha\over 16\pi^2 \fpid \mpic}\left[
{156\over45}+{\mpid\over16\pi^2\fpid}\left(12(a_2^r-2 b^r) + X_{2-}(\mu)\right)
\right]\nonumber\\
&&  (\alpha_2+\beta_2)_{\pi_0}={\alpha\over (16\pi^2 \fpid)^2\mpi }\left[
-{5009\over27} +{13435\pi^2\over720}+{16\over45}\bar{l}_2
\right]\ .
\ena
Based on resonance model estimates for the $O(p^6)$ couplings, 
numerical values for the polarizability differences at $O(p^6)$ 
were obtained~\cite{gasserivan05}. We display them below together
with the $O(p^4)$ values, which illustrates that $O(p^6)$ effects can
be rather large for these observables:
\be
\ba{cccl}
\ &  O(p^4) & O(p^6) &\    \\[1mm]
(\alpha_1-\beta_1)_{\pi_0}=& -1.0 ,&  -1.9 & 10^{-4}\,\hbox{fm}^3 \\
(\alpha_2-\beta_2)_{\pi_0}=& 20.7,& 37.6 & 10^{-4}\,\hbox{fm}^5 \ .\\
\ea
\en

The reliability of naive resonance saturation models has not
been established for $O(p^6)$ couplings.
Here, we will only make use of a model independent estimate for the 
single coupling $c^r_{34}$. This estimate is based 
on a chiral sum rule~\cite{maltmanwolfe,durrkambor} associated
with differences of correlators of two vector currents:
$\braque{V^3V^3-V^8V^8}$ or $\braque{V^{ud} V^{du}-V^{us} V^{su}}$, from which the
$SU(3)$ coupling $C_{61}^r$ can be determined.  
Using three-flavour ChPT and matching to two-flavour ChPT it can
be turned into an evaluation of the coupling $c^r_{34}$ 
(a simplified version of this sum rule was used earlier 
in ref.~\cite{knechtetnous}). 
Such matching relations have been obtained
recently  by  Gasser et  al.~\cite{gasserhaefeli}  for the  Lagrangian
operators which do not vanish  in the limit $m_u=m_d=0$. We can obtain
the  matching relation  for $c^r_{34}$  by going  to this  limit after
taking a derivative with respect to $m_u+m_d$. 
Using  the   $O(p^6)$  ChPT  calculations   of  the  $\braque{V^3V^3}$
correlator  performed   in  ref.~\cite{amorbijtal},  and   taking  the
derivative with respect to $\mpid$ one finds the following matching formula,
\bea\lbl{c34match}
&& c_{34}^r = {F^2\over 192\times 16\pi^2 \mbkd} + C_{61}^r +2 C_{62}^r
-{1\over4\times 32\pi^2}\left(\log{\mbkd\over\mu^2}+1\right)(L_9^r+L_{10}^r)
\nonumber\\
&& +O(\mbkd) .
\ena
(where $\mbkd=\lim_{m_u=m_d=0}\mkd$).
This relation allows one to evaluate  $c_{34}^r$ provided we 
further invoke a large
$N_c$ argument which implies that $C_{62}^r(\mu)$ should be suppressed
compared with  $C_{61}^r(\mu)$ when $\mu\simeq 1$ GeV.  The authors of
ref.~\cite{durrkambor} have determined that
\be
\Pi_{V^3}(0)-\Pi_{V^K}(0)= (1.92\pm0.27)\,10^{-2}
\en
(using the notations of ref.~\cite{amorbijtal} for the $\braque{VV}$ 
correlation functions). Using further the explicit chiral formulas
from     this      work~\cite{amorbijtal}     and     the     matching
relation~\rf{c34match} above, we find 
\be
c_{34}^r(\mu=m_\rho)= (1.19\pm 0.43)\,10^{-5}\ .
\en
Let us note 
the relation between $c^r_{34}$ and the combinations which appear in the
$\pi^0\pi^0$ amplitudes
\be\lbl{c34constraint}
(a_1^r+8b^r)+2(a_2^r-2b^r)= 4094\pi^4 c_{34}^r\equiv c_{34}^{eff}(\mu)\ .
\en
In other terms, the combination $6(\alpha_1-\beta_1)_{\pi^0} +\mpid
(\alpha_2-\beta_2)_{\pi^0}$ depends on the single $O(p^6)$ coupling,
$c_{34}$. Numerically, using $\mu=m_\rho$ in the ChPT expressions, the
following  relation  is obtained  between  the  dipole and  quadrupole
polarizabilities 
\be\lbl{c34relat}
 6(\alpha_1-\beta_1)_{\pi^0} + {m^2_{\pi^0}} (\alpha_2-\beta_2)_{\pi^0} =
\left(6.20 + 0.25\,c_{34}^{eff}(m_\rho)\right)\,10^{-4}\,\hbox{fm}^3
\en
with $c_{34}^{eff}(m_\rho)=4.75\pm 1.71$. 

\vskip0.2cm
\noindent{\bf b) $\pi^+\pi^-$:}\\
In the case of the charged pion, the dipole polarizabilities are given
in  terms  of  $\bar{l_6}-\bar{l_5}$  at $O(p^4)$  and  involve  three
combinations    of    couplings   $\tilde{a}_1^r$,    $\tilde{a}_2^r$,
$\tilde{b}^r$ at $O(p^6)$
\bea
&& (\alpha_1-\beta_1)_{\pi^+}= {\alpha\over 16\pi^2 \fpid \mpi}\left[
{2\over3}(\bar{l_6}-\bar{l_5}) 
+{\mpid\over16\pi^2   \fpid}\left(    \tilde{a}_1^r   +8
\tilde{b}^r + \tilde{X}_{1-}(\mu)\right)
\right]\nonumber\\
&&  (\alpha_1+\beta_1)_{\pi^+}={\alpha\mpi\over (16\pi^2 \fpid)^2 }\left[ 
8 \tilde{b}^r + \tilde{X}_{1+}(\mu) \right]\ .
\ena
The fully explicit expressions can be found in 
refs.~\cite{burgi,gasserivan06}. The quadrupole polarizabilities are 
given as follows,
\bea
&& (\alpha_2-\beta_2)_{\pi^+}= {\alpha\over 16\pi^2 \fpid \mpic}\left[
2+{\mpid\over16\pi^2\fpid}\left(12(\tilde{a}_2^r-2 \tilde{b}^r) + 
\tilde{X}_{2-}(\mu)\right) \right]\nonumber\\
&&  (\alpha_2+\beta_2)_{\pi^+}={\alpha\over (16\pi^2 \fpid)^2\mpi }\left[
-{2062\over27} +{10817\pi^2\over1440}+{8\over45}\bar{l}_1+{8\over15}\bar{l}_2
\right]\ .
\ena
Unfortunately, for the charged pion, there is no known
model independent information\footnote{A soft pion theorem due to 
Terazawa~\cite{terazawa} has sometimes been applied to the 
$\gamma\gamma\to\pi^+\pi^-$  amplitude. The theorem,  however, applies
to  the amplitude $\gamma^*(q)\gamma^*(-q)\to  \pi^+(0)\pi^-(0)$ which
is unrelated to $\gamma\gamma\to\pi^+\pi^-$: it involves different 
chiral coupling constants.}
on either of the three combinations 
$\tilde{a}_1^r$, $\tilde{a}_2^r$, $\tilde{b}^r$. In this case, we will
accept   the   estimates    of   ref.~\cite{gasserivan06}  stating that   the
polarizability difference should lie in the range 
$(\alpha_1-\beta_1)_{\pi^+}\in [4.70,6.70]\,10^{-4}\,\hbox{fm}^3$.

\section{Some details on the calculations}\lblsec{details}
\subsection{Inputs for the $\pi\pi$ T-matrix}
We describe here our inputs for $\pi\pi$ scattering amplitudes which are
needed for the $S$ and the $D$ waves. At medium energies
$\pi\pi\to\pi\pi$ phase-shifts and inelasticities have been measured in
production experiments (see \cite{martinmorganbook} for a
review). Considerable progress has been achieved recently in measuring
phase-shifts at very low energies as well as the $I=0,2$ $S$-wave
scattering lengths~\cite{NA48/2Kl4,NA48/2cusp,DIRAC}. In our analysis,
the low-energy region is emphasized in the integrals by the $s=0$
subtractions. Interpolating between the medium and low energy regions
is controlled by the set of Roy equations~\cite{Roy}.  

\begin{figure}[hbt]
\bc
\includegraphics[width=0.7\linewidth]{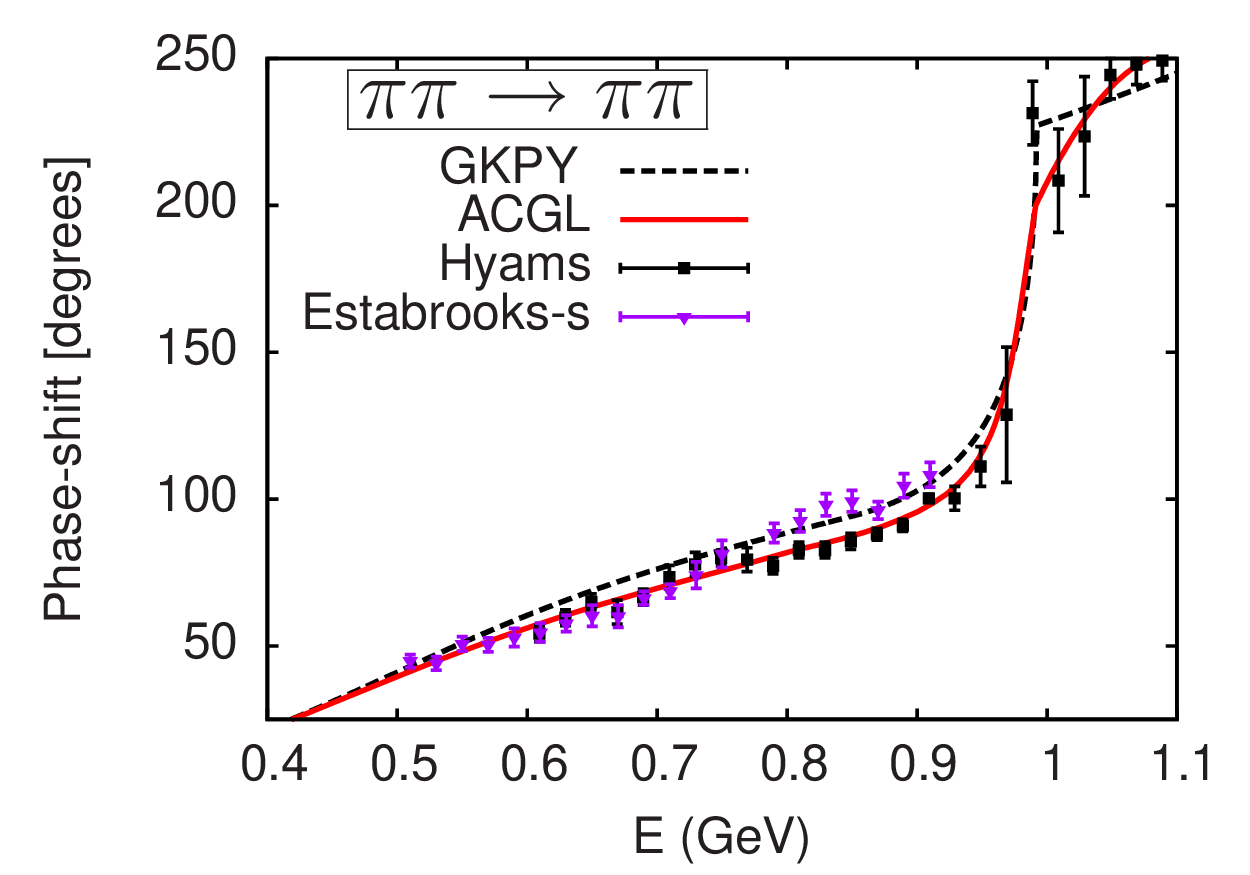}
\caption{\sl $S$-wave $I=0$ $\pi\pi$ phase-shifts below 1 GeV:
  comparison of two determinations\cite{ACGL,GKPY} used in our analysis.}
\label{fig:compruben}
\ec
\end{figure}

\noindent{\bf I=0 }\\ 
Let us consider the $S$-wave at first. 
As our main input, we will use  the
Roy equations results from ref.~\cite{ACGL} 
for the $I=0$ $S$-waves in the energy range $E\le 0.8$ GeV.  In the
range $E > 0.8$ GeV we perform fits to the experimental results of
ref.~\cite{hyams73}\footnote{It   is  likely   that   the  phase-shift
  determinations   in  the   region  $E\gapprox$   1.5  GeV   must  be
  updated~\cite{Buggzou,Lesniak}. This region plays a minor role in
  our analysis}.  
Below 1 GeV, we will also make use of the phase-shift
determinations of refs~\cite{KPY,GKPY}. These also use the Roy
equations as well as other dispersion relations as constraints. The
phase-shifts differs from~\cite{ACGL} by a few degrees in the matching
point region $E\simeq0.8$ GeV: see fig~\fig{compruben}. The
differences in the results will serve us in estimating the errors.  

A well known feature of $\pi\pi$
scattering with $I=J=0$ is the sharp onset of inelasticity when the energy 
passes the $K\bar{K}$ threshold, which is caused by the $f_0(980)$ resonance. 
In the energy range which interest us here, it is a good approximation
to ignore other inelastic channels and implement 
exact two-channel unitarity. We can deduce the required T-matrix elements
directly from experimental inputs on
$\pi\pi\to \pi\pi$ and    $\pi\pi\to K\bar{K}$ scattering
in the physical region and make use of analyticity in the unphysical region.
Also, we work  in the isospin limit and assume that  this limit can be
taken smoothly near the $K\bar{K}$ threshold. Then, 
the phase of the $T$-matrix element $T_{12}(s)\equiv g_0^0(s)$ 
should be equal to the phase of
$T_{11}(s)\equiv   f_0^0(s)$ (i.e.   to   the   elastic   $\pi\pi$
phase-shift) when $\sqrt{s}=2m_K$. 
This threshold phase is actually not very well known at present:
for illustration, the $K$-matrix fit of Hyams et al.~\cite{hyams73}
gives   $\delta(2m_K)\simeq  175^\circ$  while   the  GKPY~\cite{GKPY}
analysis gives $\delta(2m_K)\simeq  227^\circ$. 
Concerning $T_{12}$, the two experiments
by Cohen et al.~\cite{cohen} and Etkin et al.~\cite{etkin} 
are in disagreement close to the 
threshold, for both the phase and the modulus. The results of Cohen et
al. are in accord with a large value of the threshold phase 
$\delta(2m_K)\simeq  220^\circ$  while,  on  the  contrary,  Etkin  et
al.  favour   a  value   smaller  than  $150^\circ$.   An  alternative
possibility associated with the results of  Etkin et al., is that of a
fast variation of the phase in between the $K^0\bar{K}^0$ and the 
$K^+\bar{K}^-$  thresholds, as suggested  in ref.~\cite{AMP}.  In that
case, the isospin limit would not be smooth.

\begin{figure}
\bc
\includegraphics[width=0.6\linewidth]{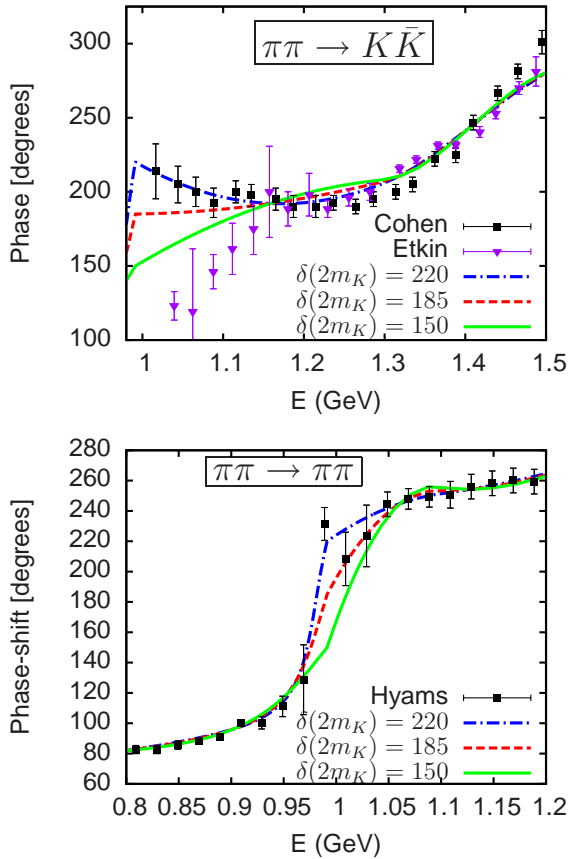}
\caption{\sl Fits to the $\pi\pi\to K\bar{K}$ phase (upper figure) and
  the $\pi\pi$ phase-shifts corresponding to different input values of
  the $K\bar{K}$ threshold phase.}
\label{fig:phase12}
\ec
\end{figure}

In order  to probe the sensitivity  of the $\gamma\gamma$  data to the
value of the threshold phase, we have performed our fits 
allowing it to vary in the range $150^\circ \le \delta({2m_K})  \le 220^\circ$. 
Figure~\fig{phase12} (upper plot)
shows our fits of the $\pi\pi\to K\bar{K}$ phase compared to 
the two data sets. 
Below the inelastic threshold, 
in the region $0.8\ \hbox{GeV}\le E \le 2m_K$ we adopt the following simple 
description of the $\pi\pi$ phase-shift, involving a Breit-Wigner term
plus a linear background
\be\lbl{simplepar}
\delta_{\pi\pi}(\sqrt{s})=   a  +   b  s   +   \arctan{m_{0}  \Gamma_{0}\over
  m^2_{0}-s}\ ,
\qquad 0.8\ \hbox{GeV}\le \sqrt{s}\le 2m_K\ .
\en
Assuming given values for the phase-shift at $E=0.8$ GeV and $E=2m_K$ fixes
the parameters $a$, $b$ in terms of $m_{0}$, $\Gamma_{0}$. These
two parameters are then fitted to the experimental data in this region. 
A few values corresponding to different inputs for the threshold phase
$\delta_{\pi\pi}(2m_K)$ are collected in table~\Table{m0gam0}. The mass and
width of the $f_0(980)$ resonance in this simple parametrization are in
reasonable agreement with the PDG values. The width is seen to be
rather sensitive to the input threshold phase.
The corresponding curves are  shown  on  fig.~\fig{phase12}.  
At  energies  above $2m_K$,  we  describe  both  the  $\pi\pi\to
\pi\pi$  and  $\pi\pi\to K\bar{K}$ phase-shifts by piecewise 
polynomial functions fitted to experiment. 
\begin{table}[ht]
\bc\bt{|l| c c c c |}\hline
\TT$\delta_{\pi\pi}(2m_K)$ & 180$^\circ$ &200$^\circ$ &210$^\circ$ &220$^\circ$ \\ \hline
$m_0$ (GeV) &0.987 & 0.984 & 0.983 & 0.981 \\
$\Gamma_0$ (GeV) & 0.056 & 0.039 & 0.033 & 0.028 \\ \hline
\et
\caption{\sl Mass and width (in GeV) of the $f_0(980)$ resonance
  arising from fitting the parametrization~\rf{simplepar} to the data
  in the range $0.8\ \hbox{GeV}\le E\le 2m_K$ with different threshold
  phase inputs.} 
\lbltab{m0gam0}
\ec
\end{table} 

The modulus of $T_{12}(s)$ 
it needed in the unitarity equations~\rf{pwunitar00} in the unphysical
region  $4m^2_\pi \le  s \le  4m^2_K$. Using  analyticity  and elastic
unitarity it can  be determined by the MO  method. Since the left-hand
cut  of $T_{12}(s)$  can be  expressed  in terms  of $\pi  K\to \pi  K$
phase-shifts this MO equation is actually one component of the set of coupled
Roy-Steiner  equations (see  e.g.~\cite{BDM}). We  will  employ
here a simplified but reasonably accurate representation,
\be\lbl{T12omnes}
T_{12}(s)= \left(A_0 +s(A +Bs+Cs^2)\right)\exp\left[
  {s\over\pi}\int_{4\mpid}^\infty 
ds' {\delta_{12}(s')\over s'(s'-s)} \right]
\en
where the two  parameters $A$ and $A_0$ are chosen  so as to reproduce
the      values     $T_{12}(0)=0.097$      and      its     derivative
$\dot{T}_{12}(0)=1.126$ $\hbox{GeV}^{-1}$ 
obtained from  the full Roy-Steiner  equations~\cite{BDM} and the
two remaining parameters $B$ and $C$ are fitted to the experimental data in the
range $[1-1.5]$ GeV. 

\begin{figure}[hbt]
\bc
\includegraphics[width=12cm]{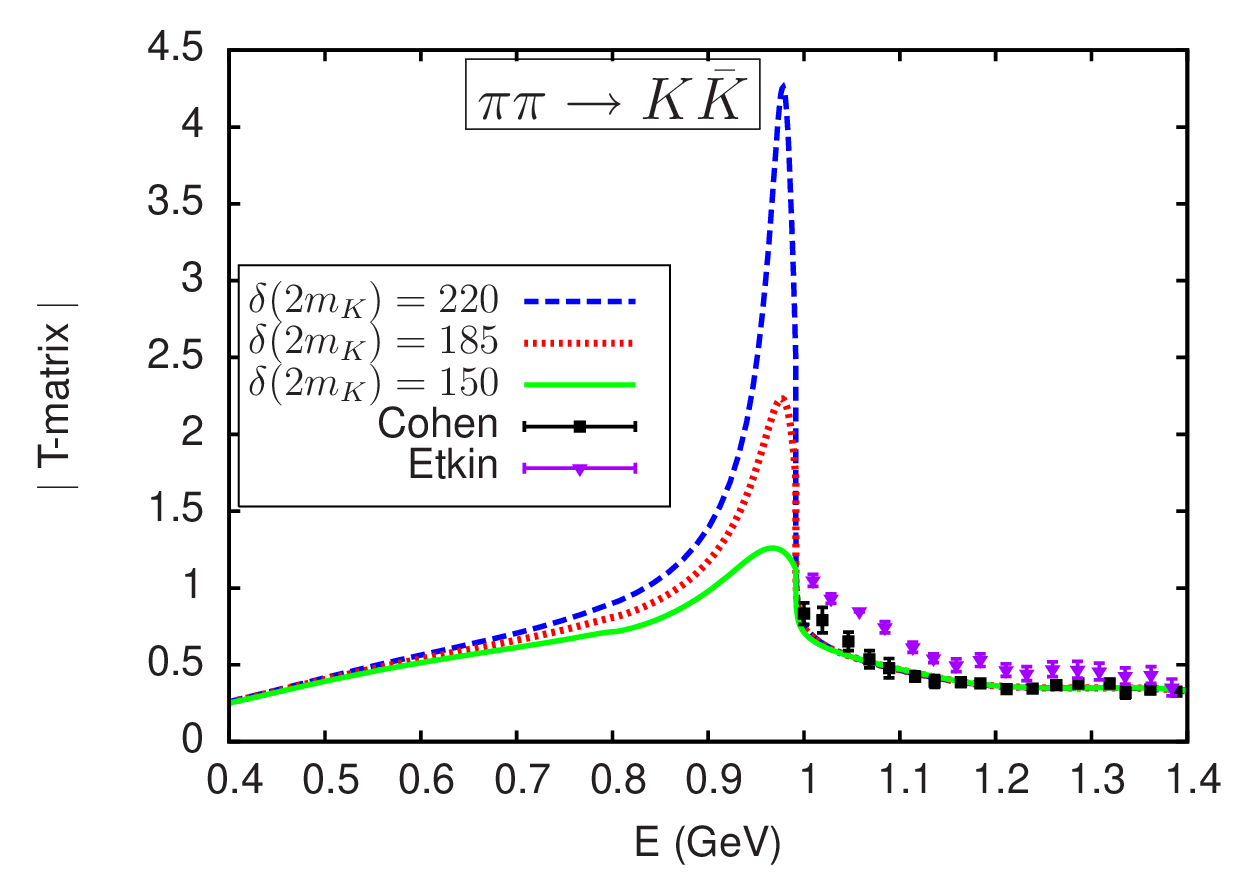}
\caption{\sl Modulus of the $\pi\pi\to K\bar{K}$ $S$-wave $T$-matrix element 
using the representation~\rf{T12omnes} fitted to the data, 
for several values of
the $K\bar{K}$ threshold phase-shift.}
\label{fig:figtmat12}
\ec
\end{figure}

The modulus of $T_{12}$ displays a peak associated with the
$f_0(980)$ resonance. As one can expect from the formula~\rf{T12omnes}, 
the size of this peak is strongly correlated 
with the value of the threshold phase $\delta_{12}(m_K)$ . 
This is illustrated in fig.~\fig{figtmat12}.
The incompatibility between the results of Cohen et al. and of Etkin et al.
below $E\simeq 1.2$ GeV is also apparent on this figure. Under the assumption
of two channel unitarity, $T_{12}$ is related to the inelasticity parameter
$\eta_{\pi\pi}$ in $\pi\pi$ scattering by
\be
T_{12}(s)= {\sqrt{1-\eta^2(s)}\over2\sqrt{\sigma_\pi(s)\sigma_K(s)}}
\en
with $\sigma_P(s)=\sqrt{1-4m_P^2/s}$. 
The results of Hyams et al.~\cite{hyams73} on the $\pi\pi$ inelasticity 
which have relatively large error bars are compatible with both
refs.~\cite{cohen,etkin}. 
Once   the  $2\times2$  $T$-matrix   is  defined,   the  corresponding
$2\times2$ Omn\`es matrix can be computed numerically (see~\cite{dgl,moussNf}
for details). 

For the $D$-wave, we have relied on the two-channel $K$-matrix representation
of Hyams et al.~\cite{hyams73} for representing the $\pi\pi$ phase-shift,
updating the mass and width of the $f_2(1270)$ resonance to the PDG values. 
In this case the $K\bar{K}$ inelastic channel is not the
physically dominant one below 1.3 GeV but is used
as an effective description of inelasticity.
Furthermore, the inelasticity quoted by the PDG at the energy of
the  $f_2(1270)$ is  significantly smaller  than the  one  obtained by
Hyams et al. 
In practice, we have used a one-channel Omn\`es representation 
using the $T$-matrix phase instead of the $S$-matrix phase in the inelastic
region.

\begin{figure}
\bc
\includegraphics[width=12cm]{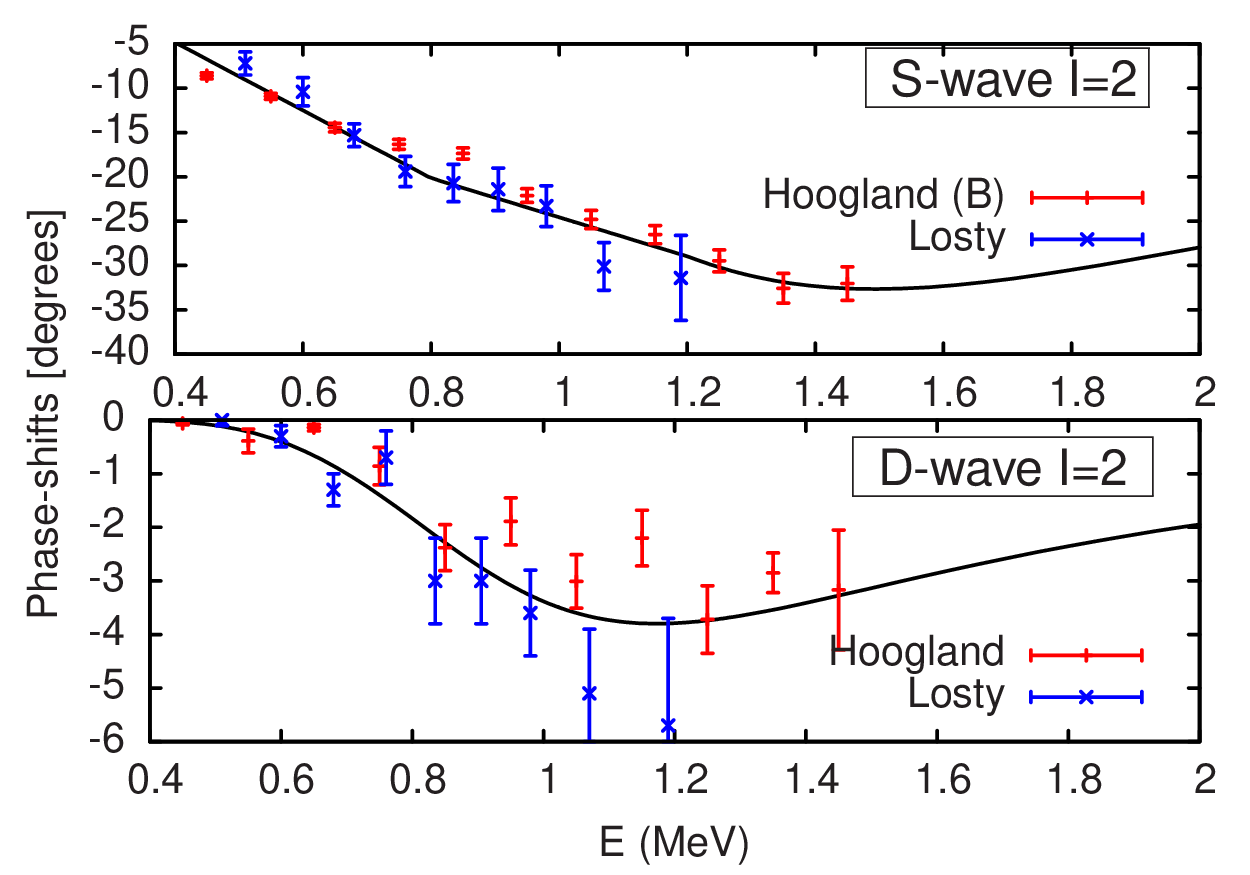}
\caption{\sl $\pi\pi\to \pi\pi$ isospin 2 phase-shifts: $S$-wave (upper curve)
and $D$-wave (lower curve).}
\label{fig:isospin2}
\ec
\end{figure}

\noindent{\bf I=2:}\\ 
In this case we have ignored inelasticity. For the $S$-wave, we use the Roy
parame\-trization of ref.~\cite{ACGL} below 0.8 GeV and make a simple
fit to the data of refs.~\cite{losty,hoogland} at higher energy. For the
$D$ wave we make a simple fit to the data found in the same references
in the whole energy region. These fits are shown in fig.~\fig{isospin2}

\subsection{Experimental data}
The main body of experimental data which we used are described in the 
publications~\cite{bellemori2pic_1,bellemori2pic_2} (charged pions) and
\cite{belleuehara2pi0_1,belleuehara2pi0_2} (neutral pions) by the Belle
collaboration. They have measured differential cross-sections in the
range $\cos\theta\le 0.6$ for $\pi^+\pi^-$ and $\cos\theta\le 0.8$
for $\pi^0\pi^0$. We have taken into account all their data in the
energy range $E\le 1.28$ GeV i.e. 520 data points for $\pi^0\pi^0$ and 
$1152$ data points for $\pi^+\pi^-$. In addition we have taken into account
earlier experimental measurements of cross sections integrated over 
$\cos\theta$ (in the same ranges as indicated above) from the Crystall Ball
collaboration~\cite{crystalball} ($\pi^0\pi^0$) as well as MarkII and Cello
~\cite{markII,cello} ($\pi^+\pi^-$). We have assigned equal weights to all
the data points. Obviously then, the $\chi^2$ is completely dominated
by Belle's results.

\subsection{Parameters to be fitted}
The dispersive representations for the partial-wave amplitudes as written in 
sec.~\sect{omrepres} involve 10 subtraction parameters. 
Not all of them will be determined from the fit. \\
1) For the I=2 $D$-waves we have actually assumed an unsubtracted
dispersion relation (i.e. the corresponding parameters $c^{(2)}$ 
and $d^{(2)}$ are determined from sum rules, see eq.~\rf{cIdIsr}).\\
2) We ave used some chiral constraints. Firstly, 
we have fixed the parameter 
$b^{(0)}_K$ to be equal to its ChPT expression at one loop
\be
b^{(0)}_K = -{( L^r_9+L^r_{10})\over \fpid} + O(p^6)\ .
\en 
Taking $L_9$ and $L_{10}$ from table 2 of ref.~\cite{ecker_revue} gives
\be
b^{(0)}_K \simeq -0.40\pm 0.30\ .
\en
As a further constraint we use the known value of the $O(p^6)$ 
coupling  $c_{34}^r$, as  discussed  in sec.~\sect{constraintch}.  This
provides one relation among the seven remaining parameters and leaves six
parameters to be fitted. Finally, 
we perform the fits also imposing that the dipole polarizability
difference of the charged pion lies in the range allowed by 
ChPT~\cite{gasserivan06}.
\section{Results of the fit}
\begin{figure} 
\bc
\includegraphics[width=\linewidth]{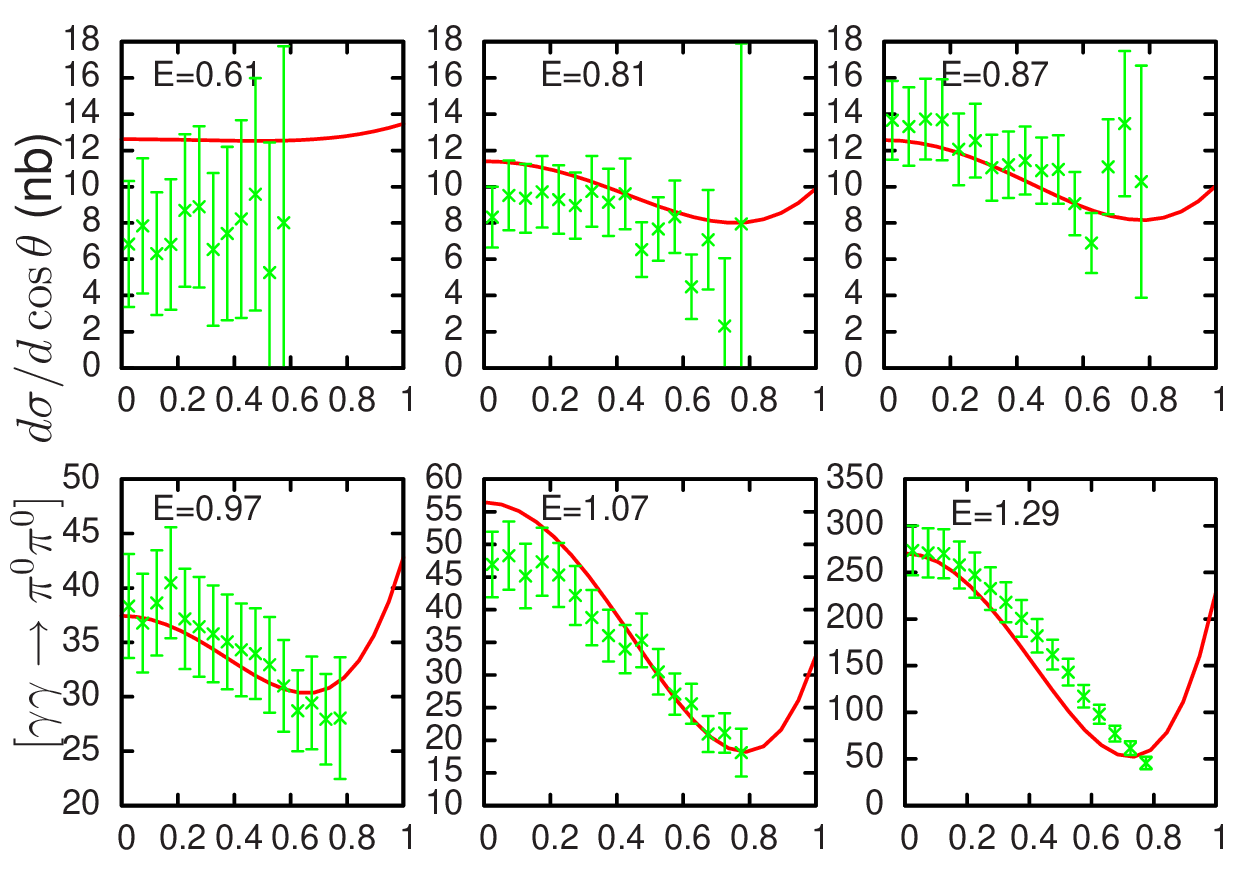}
\includegraphics[width=0.745\linewidth]{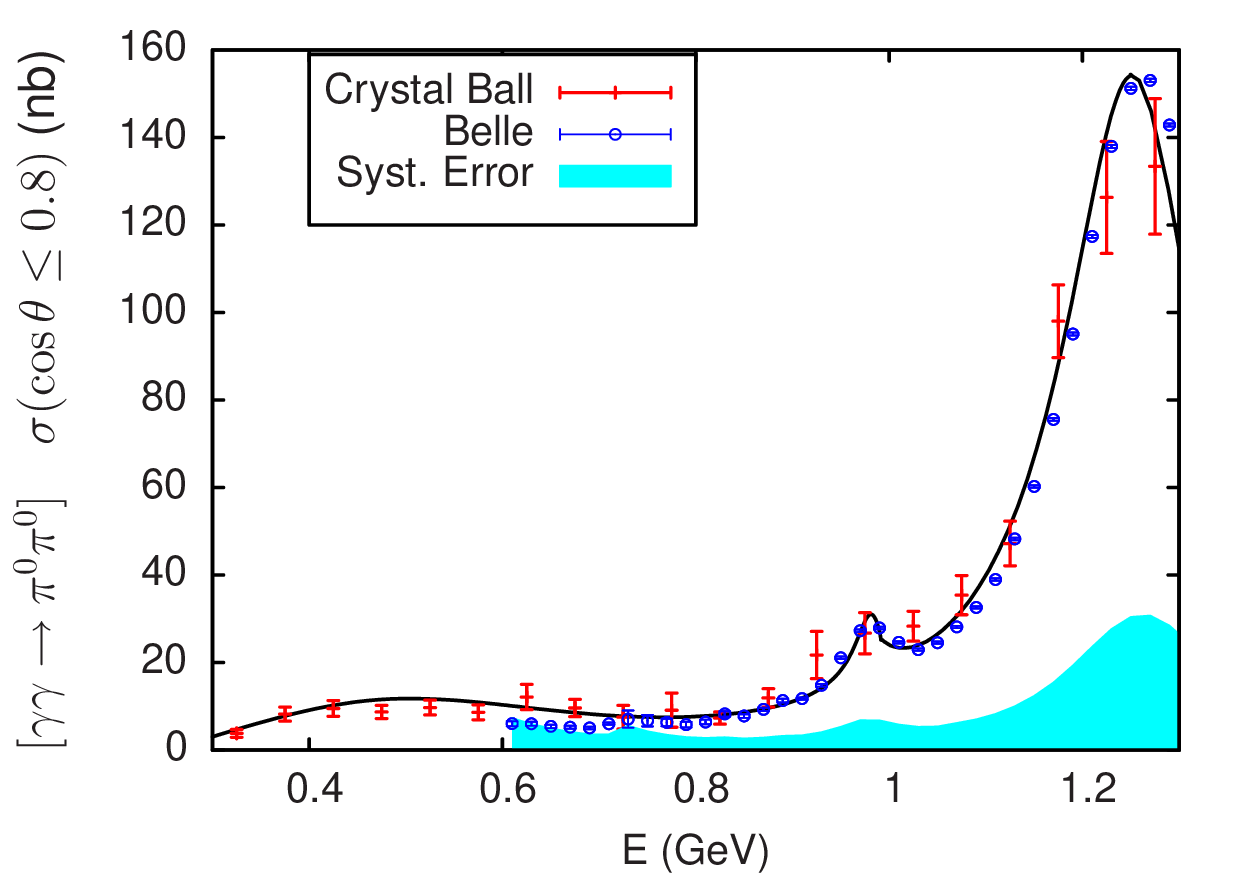}
\caption{\sl Differential cross-sections (systematic and statistical
  errors added in 
quadrature)  and integrated  cross-section (Belle's  systematic errors
are shown separately) 
for $\gamma\gamma\to \pi^0\pi^0$. The solid line is the result of our
constrained fit.}
\label{fig:secspi0pi0}
\ec
\end{figure}

\begin{figure} 
\bc
\includegraphics[width=\linewidth]{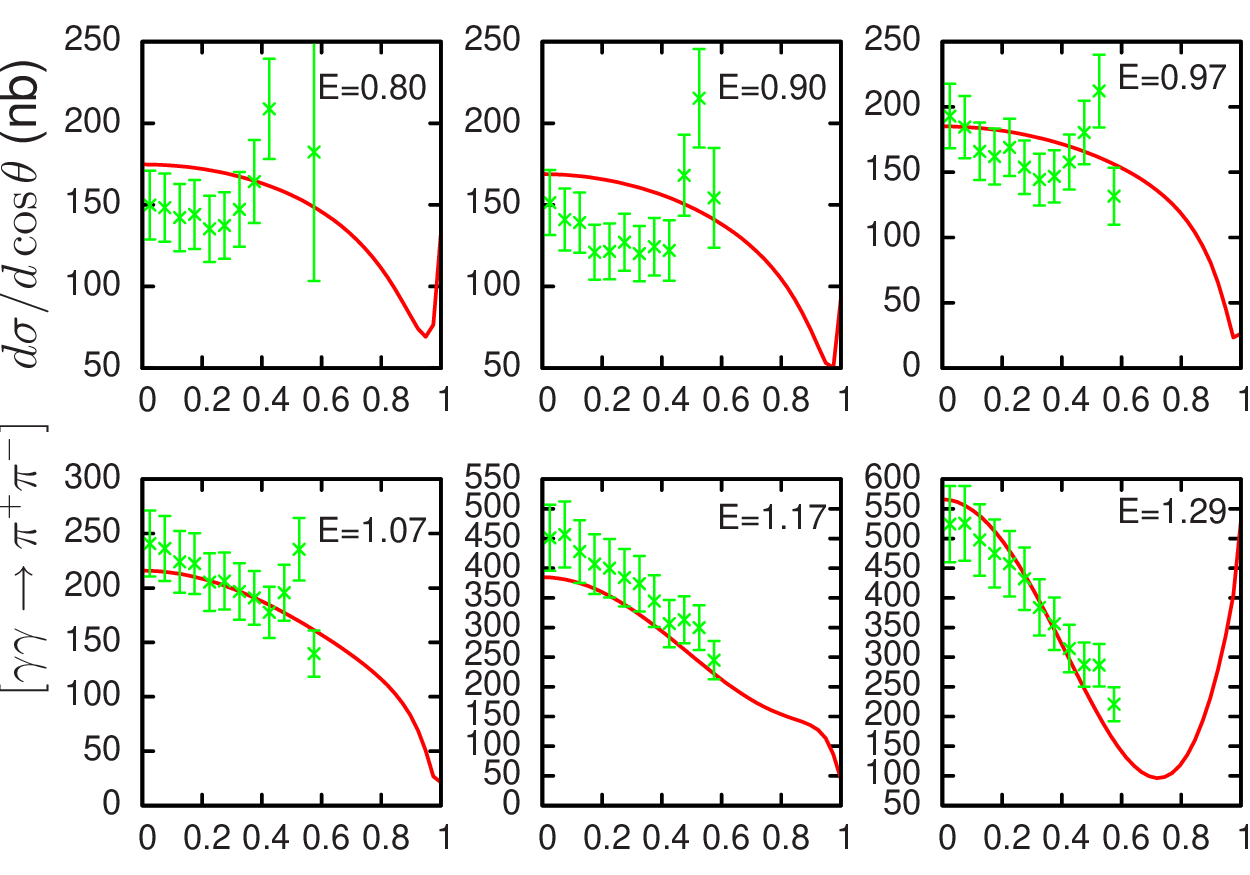}
\includegraphics[width=0.745\linewidth]{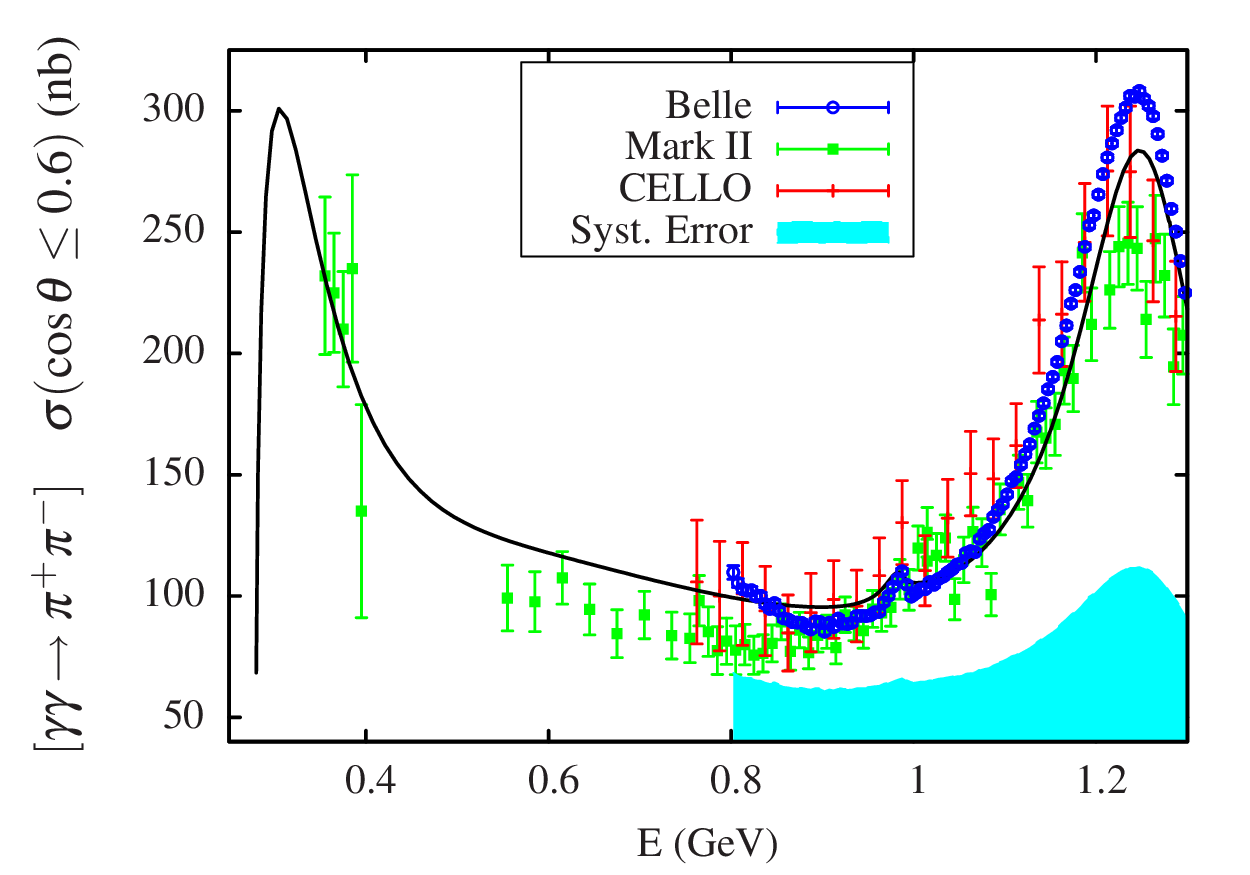}
\caption{\sl Same as fig.~\fig{secspi0pi0} for $\gamma\gamma\to \pi^+\pi^-$}
\label{fig:secspi+pi-}
\ec
\end{figure}

\begin{figure}[t]
\bc
\includegraphics[width=0.7\linewidth]{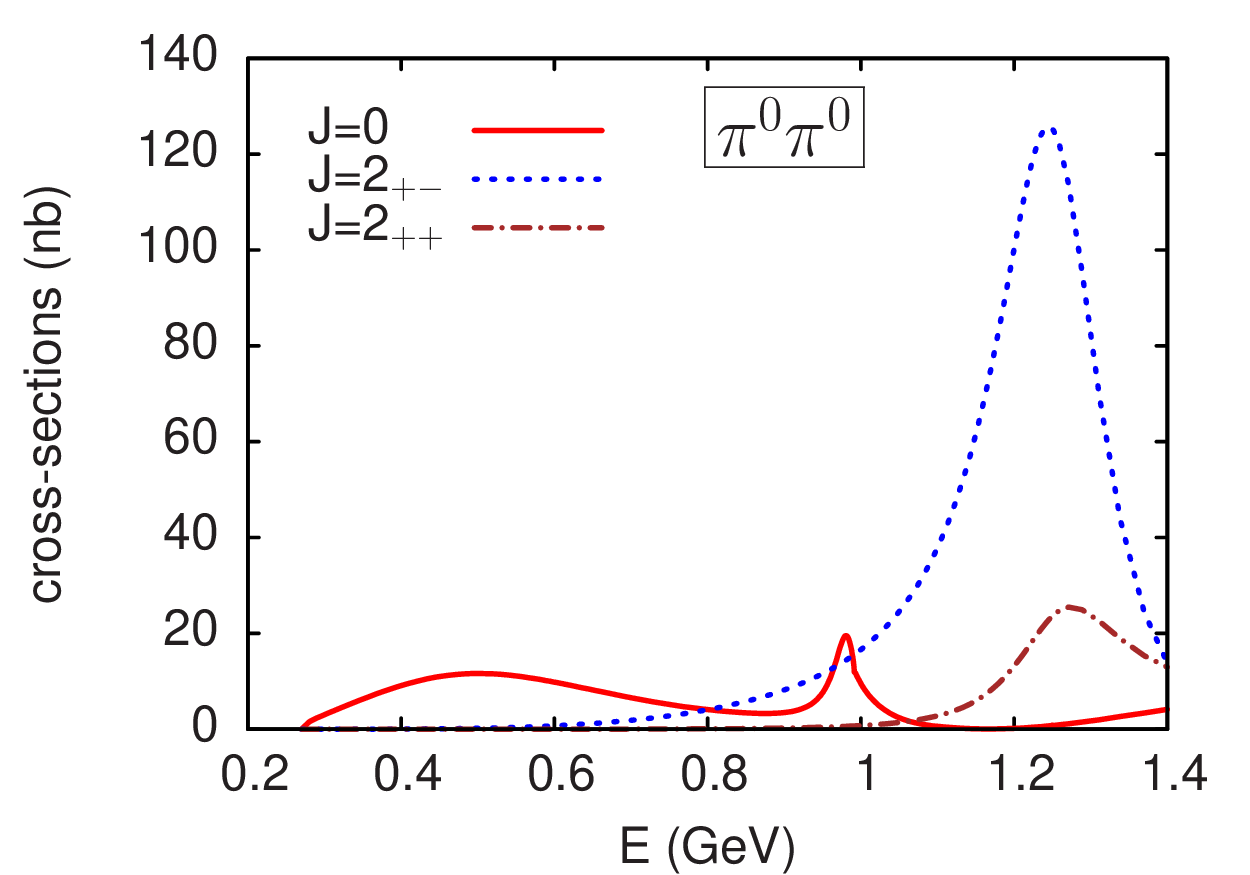}
\caption{\sl Contributions  from the $S$ and the  $D$ partial-waves to
  the $\gamma\gamma\to \pi^0\pi^0$ integrated cross-section}
\label{fig:partwav}
\ec
\end{figure}

\begin{table}
\bc
\bt{|l|r|r|}\hline
\  & $N_{data}$ &  $\chi^2/ N_{data}$  \\ \hline
$\pi^0\pi^0$ Belle        & 520 &  1.26 \\
$\pi^0\pi^0$ Crystal Ball &  21 &  1.13 \\ \hline
$\pi^+\pi^-$ Belle        &1152 &  1.39 \\
$\pi^+\pi^-$ Cello        &  23 &  1.23 \\
$\pi^+\pi^-$ Mark II      &  67 &  2.85 \\ \hline
\et
\caption{\sl $\chi^2$ results from the constrained six parameters fit}
\lbltab{chi2}
\ec
\end{table}
Differential cross-sections for $\gamma\gamma\to\pi^0\pi^0$ and 
$\gamma\gamma\to\pi^+\pi^-$ are evaluated with the $J=0$ and the $J=2$
amplitudes          computed          as         explained          in
secs.~\sect{omrepres},~\sect{leftcut},~\sect{details}.              For
$\gamma\gamma\to\pi^+\pi^-$  all the $J\ge4$  amplitudes corresponding
to the Born term are also included.
The values of the $\chi^2$ results after minimization, corresponding to the
various data sets, and adding the statistical and the systematic errors
in quadrature, are shown in table~\Table{chi2}. These numbers correspond
to  a choice of  left-cut cutoff  $\Lambda=-5$ GeV$^2$  and $K\bar{K}$
threshold phase 
$\delta(2m_K)=200^\circ$. The values of the $\chi^2$ for the various
data sets are similar (which indicates compatibility)
with the exception of the Mark II data~\cite{markII} which shows some
deviation.  This feature was observed also
in some previous analysis~\cite{pennington+belle}.
In more detail, the ability of our constrained dispersive representations
to reproduce the experimental data is illustrated on fig.~\fig{secspi0pi0} 
(for $\pi^0\pi^0$) and fig.~\fig{secspi+pi-} (for $\pi^+\pi^-$). 
The $\pi^0\pi^0$ amplitude is somewhat simpler than the $\pi^+\pi^-$ one
due to the absence of the  direct Born contribution in that case. This
can be seen from the shapes of the differential cross-sections. In the
energy region under consideration here  only the $S$-wave and two $D$-waves
effectively    contribute   to    the    $\gamma\gamma\to \pi^0\pi^0$
amplitude.  In the  energy  region of  the  $f_2(1280)$ a  significant
contribution from  the $S$-wave  background is needed  for reproducing
the experimental cross-section.  Our $S$-wave amplitude based on two-channel
unitarity    cannot     be    quantitatively    trusted     in    this
region.  Figure~\fig{partwav}  illustrates  the  role of  the  various
partial-waves in the integrated cross section.

\begin{table}[h]
\bc
\bt{|l||l|l|l|l|l|}\hline
\ & $b^{'(0)}$ & $b^{'(0)}_K$ & $b^{'(2)}$ & $c^0$ & $d^0$ \\ \hline 
Fit              &$-7.99$ & $-7.16$ & $3.47$ & $0.22$ & $-0.48$ \\  \hline 
SR ($\Lambda=-5$)&$-6.00$ & $-5.61$ & $3.29$ & $0.19$ & $-0.54$ \\  
SR ($\Lambda=-3$)&$-7.62$ & $-6.84$ & $3.22$ & $0.10$ & $-0.57$ \\  \hline 
\et
\caption{\sl Comparison of the subtraction parameters as determined from
a fit of the experimental data and as determined from sum rules, as a function
of the cutoff $\Lambda$ on the left-cut integration. Units are in
appropriate powers of GeV.}
\lbltab{fitparams}
\ec
\end{table}
The dispersive representations are based on over-subtracted dispersion
relations. Assuming reasonable high-energy behaviour,
five of the subtraction parameters could be written
as sum  rules (see sec.~\sect{sumrules}).  In practice, the  result of
such sum rules depend on the 
cutoff $\Lambda$ that one introduces on the left-hand cut integration
(since the precise behaviour of the integrand 
for large negative $s$ is not known). 
Table~\Table{fitparams} shows the numerical (central) 
values of these parameters, as generated by the fit, and the sum rule
evaluations. The table shows that, for physically reasonable values
of the cutoff, $\Lambda\simeq -3,-5$ GeV$^2$, the fitted values are  
qualitatively in agreement with the sum rule ones.

 \begin{figure}[htb] 
 \bc
 \includegraphics[width=0.6\linewidth]{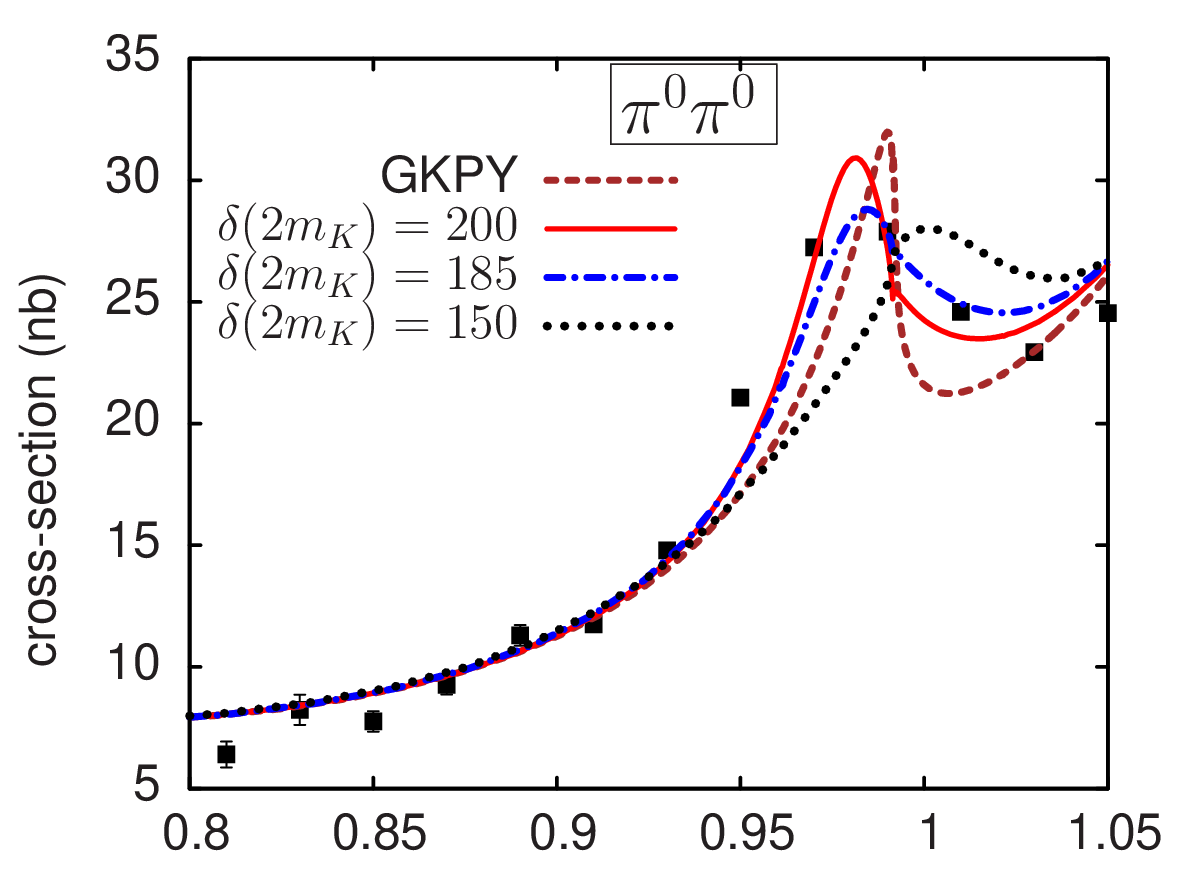}\\
 \includegraphics[width=0.6\linewidth]{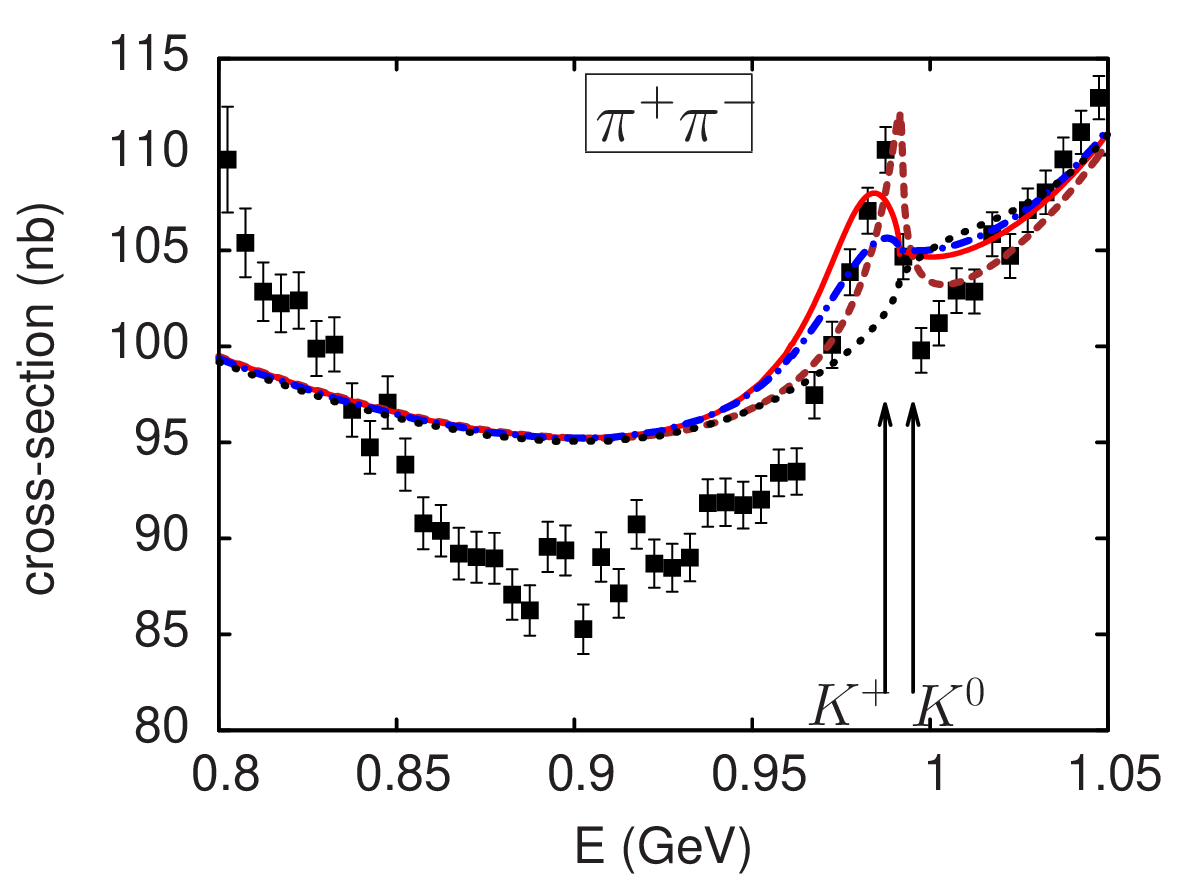}
 \caption{\sl   Integrated  cross-sections  in   the  region   of  the
 $f_0(980)$ peak. The  solid, dashed and dotted curves  are the result
 of  our  MO  representation  corresponding  to three  values  of  the
 $\pi\pi$  phase-shift at  the $K\bar{K}$  threshold.  The long-dashed
 curves correspond to using the phase-shifts of ref.~\cite{GKPY}. 
 The experimental results  of  
 Belle~\cite{belleuehara2pi0_1,bellemori2pic_1} are  
 shown with their statistical errors only. }
 \label{fig:f0980}
 \ec
 \end{figure}

\vskip2mm
\noindent{\bf $\bullet$ $f_0(980)$ region:}\\
Our parametrization of the input $\pi\pi$ phase-shifts allows for some freedom
to   vary   the  value   of   the   phase-shift   at  the   $K\bar{K}$
threshold.  While the  overall $\chi^2$  is hardly  sensitive  to this
small energy region,  Belle's data provides a detailed  picture of the
$f_0(980)$ peak because of the large statistics. 
A comparison of our results corresponding to different values of the
threshold phase-shift $\delta(2m_K)$, 
with Belle's data, is illustrated on fig.~\fig{f0980}.
For clarity of the figure, the systematic errors are not shown, in this region
they are of the order of 12 nb for $\pi^+\pi^-$ and 3.5 nb for $\pi^0\pi^0$. 
In the case of $\pi^0\pi^0$, comparison with the data 
favours values of the threshold phase $\delta(2m_K)\gapprox 180^\circ$:
for smaller values the peak is too flat and displaced to the
right. The value  $\delta(2m_K)\simeq 200^\circ$ eventually provides 
the closest agreement with the experimental peak.
The shape of the $f_0(980)$ in $\pi^0\pi^0$ and in $\pi^+\pi^-$ is
predicted to differ because of the different sign of the interference
between the $I=0$ resonant amplitude and $I=2$ amplitude. 
There is some indication of this feature in the data.
Comparing the MO results with the central values of the data for
$\pi^+\pi^-$ one must keep in mind that the systematics are larger
than for $\pi^0\pi^0$. However, essentially the same value of the   
threshold phase $\delta(2m_K)\simeq 200^\circ$ also provides the best
agreement with the shape of the $f_0(980)$ peak. 
When  $\delta(2m_K)$ gets  smaller than $\simeq 170^\circ$
the structure resembles a cusp rather than a peak. 
Belle's statistics are very high for $\pi^+\pi^-$ and the energy bins 
$\Delta E=5$ MeV are smaller than the spacing between the
$K^0\bar{K^0}$ and the $K^+{K^-}$ thresholds ( $\simeq 8$ MeV, the
two thresholds are indicated by arrows in fig.~\fig{f0980}). Belle's
data are compatible with a small isospin breaking at the $K\bar{K}$
threshold since agreement with the MO amplitude (which has no isospin
breaking) in this region is not
worse than elsewhere. A strong isospin breaking scenario has been
proposed by  Au et al.~\cite{AMP}, according to which the
$\pi\pi\to  K\bar{K}$ phase  drops  very sharply  in  between the  two
thresholds. Such a scenario is also not ruled out by Belle's data. 
Isospin    breaking   near    the   $K\bar{K}$    threshold   involves
$a_0(980)-f_0(980)$ mixing.  Studies of how this affects  the shape of
the $a_0(980)$ peak have been performed (e.g.~\cite{achasovmix,hanhart}).

 \begin{figure}[htb]
 \bc
 \includegraphics[width=0.6\linewidth]{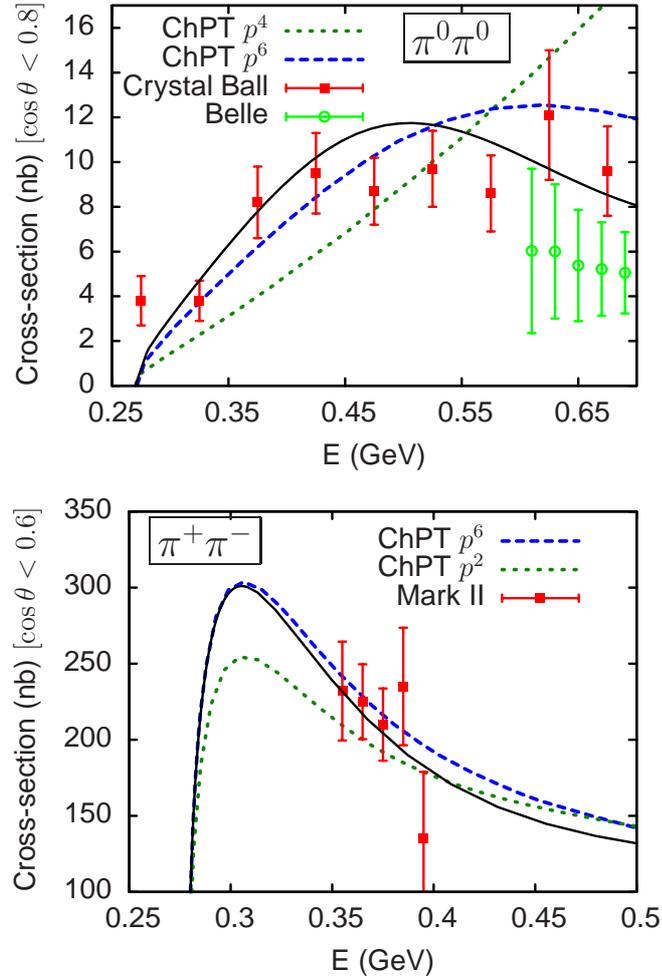}
 \caption{\sl Integrated $\gamma\gamma$ cross-sections at low energy. The 
experimental results from~\cite{crystalball,markII} are shown and also the
results from ChPT calculations as given in ~\cite{gasserivan05,gasserivan06}. 
The solid line is the result from the MO amplitudes.}
 \label{fig:lowenergy}
 \ec
 \end{figure}

\begin{figure}
\bc
\includegraphics[width=0.7\linewidth]{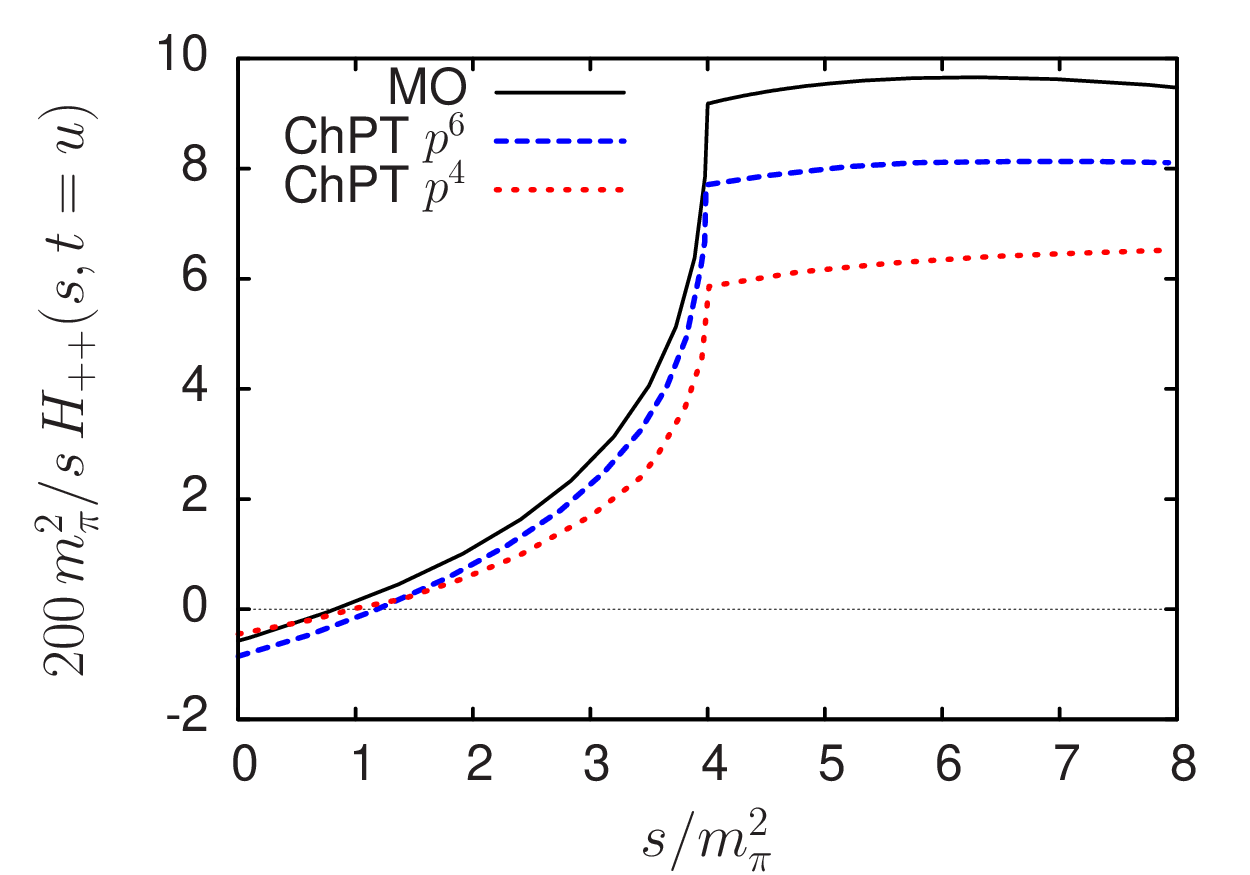}
\caption{\sl Amplitude $H_{++}(s,t=u)/s$ in the sub-threshold region
and $\vert  H_{++}(s,t=u)\vert/s$ above the threshold (as  in fig.~8 of
ref.~\cite{bellucci}) compared to ChPT results.}
\label{fig:HPP}
\ec
\end{figure}

\vskip2mm
\noindent{\bf $\bullet$ Low energy region, pion polarizabilities}\\
Next, we consider the low energy region. Fig.~\fig{lowenergy} shows
our result for the integrated cross-sections in this region and also shows, for
comparison, the result from ChPT calculations at NLO and NNLO. 
As  another  comparison,  we   show  in  fig.~\fig{HPP}  the  amplitude
$H_{++}(s)/s$ in  the sub-threshold region  normalized as in  fig.~8 of
ref.~\cite{bellucci} (above  the threshold the modulus  is shown).  In
this region, the dispersive amplitude  lies rather close to the chiral
amplitudes. In particular, it has  an Adler zero close to $s=\mpid$ as
has been anticipated in ref.~\cite{morganpenning88}. 
The  differences between  the dispersive  results and  ChPT are  to be
attributed to effects of chiral order $O(p^8)$ and also to 
some differences in $O(p^6)$ coupling constants (see below).

\begin{table}[hbt]
\bc
\bt{|c|c|c|}\hline
\ {\large$\pi^0$} & Fit & ChPT +Res. mod. \\ \hline
\TT$(\alpha_1-\beta_1)_{\pi^0}$ & $-1.25\pm0.08\pm0.15$ & $-1.9\pm0.2$ \\
\TT$(\alpha_1+\beta_1)_{\pi^0}$ & $ 1.22\pm0.12\pm0.03$ & $1.1\pm 0.3$ \\
\TT$(\alpha_2-\beta_2)_{\pi^0}$ & $ 32.1\pm0.9\pm1.9$ & $37.6\pm3.3$ \\
\TT$(\alpha_2+\beta_2)_{\pi^0}$ & $-0.19\pm0.02\pm0.01$ & $0.04$\BB \\ \hline
\ {\large$\pi^+$} & & \\ \hline
\TT $(\alpha_1-\beta_1)_{\pi^+}$ & $ 4.7$ & $5.7\pm1.0$  \\ 
\TT $(\alpha_1+\beta_1)_{\pi^+}$ & $ 0.19\pm0.09\pm0.03$ & $0.16[0.16]$ \\ 
\TT $(\alpha_2-\beta_2)_{\pi^+}$ & $14.7\pm1.5\pm1.4$ & $16.2[21.6]$ \\ 
\TT $(\alpha_2+\beta_2)_{\pi^+}$ & $ 0.11\pm0.03\pm0.01$ & $-0.001$ \BB \\ \hline
\et
\caption{\sl  Results for  dipole and  quadrupole  polarizabilities (in
  units of $10^{-4}$ fm$^3$ and $10^{-4}$ fm$^5$ respectively) of the
$\pi^0$ and $\pi^+$ compared with the
values from ChPT at $O(p^6)$ associated with a model for the LEC's
(from ref.~\cite{gasserivan05} for the $\pi^0$, 
and~\cite{gasserivan06} for the $\pi^+$. The numbers in brackets correspond 
to using the ENJL model~\cite{bijprad96} for the LEC's).
The central values shown correspond to a fit
with a  left integration cutoff  $\Lambda=-5\ GeV^2$. The  first error
corresponds to  the variation of  the input parameters and  the second
error reflects the uncertainties in the experimental $\gamma\gamma$ data.} 
\lbltab{polartable}
\ec
\end{table}

Expanding the
$\gamma\gamma$ amplitudes around $s=0$ with $t=\mpid$, one accesses
pion polarizabilities (see  sec.~\sect{polarizab} ). The results for these 
quantities deduced from our fitted amplitudes are collected in 
table~\Table{polartable} and compared with the results from ChPT at 
NNLO as given in~\cite{gasserivan05,gasserivan06}. 
Our fit was performed with the constraint that for the charged pion,
the polarizability difference $(\alpha_1-\beta_1)_{\pi^+}$ should lie
within the range of the ChPT calculation. The fit prefers the lowest value in 
the allowed range.  This trend differs from the  result of Fil'kov and
Kashevarov~\cite{filkov06} who fitted the charged amplitude only.
For the neutral pion, we have imposed a constraint between the 
dipole and the quadrupole polarizabilities (see ~\rf{c34relat}). The result
for the polarizability difference $(\alpha_1-\beta_1)_{\pi^0}$ is then
in acceptable  agreement with the ChPT prediction. Our results for the
dipole polarizability sums are in agreement with ChPT. For the quadrupole
polarizability differences, our results are slightly smaller than ChPT. 
This corresponds to somewhat different results for the chiral coupling
constant  combinations $a_1^r$,  $a_2^r$  and $b^r$  for which  simple
models have been used in ref.~\cite{gasserivan05,gasserivan06}. 
Table~\Table{compa1a2b} shows the values of these constants resulting
from the fit and compared with those from a resonance model and also from
the ENJL model~\cite{bijprad96}.  We note that, in the NJL model, 
the dipole and quadrupole polarizabilities can be calculated directly, 
e.g.~\cite{bernardvautherin,hiller}. 
Our predictions differ from $O(p^6)$ ChPT for the sums of the quadrupole 
polarizabilities. The $O(p^6)$ chiral couplings
cancel out in the expressions  of these observables. As a consequence,
they are expected to be sensitive to 
effects of chiral order $p^8$~\cite{gasserivan06}. 

Let us also mention that our results for $(\alpha_2-\beta_2)_{\pi^0,\pi^+}$
are somewhat smaller than those obtained in 
refs.~\cite{filkov05,filkov06},
\be
(\alpha_2-\beta_2)_{\pi^0}= (39.70\pm0.02)\,10^{-4} \quad \quad
(\alpha_2-\beta_2)_{\pi^+}= (25.0^{+0.8}_{-0.3})\,10^{-4}\ 
\en
(in units  of fm$^5$) from  fitting subtracted dispersive
representations (they used a combination of $s$-fixed and $t$-fixed
dispersion relations) of $\gamma\gamma\to \pi^+\pi^-$ (ref.~\cite{filkov06}),
$\gamma\gamma\to \pi^0\pi^0$ (ref.~\cite{filkov05}), 
amplitudes to experimental data.
This difference in the results is to be attributed, we believe, to our
combining $\pi^0\pi^0$ and $\pi^+\pi^-$ data in the fits as well
as our using a more sophisticated treatment of the final-state interaction
in  the  $S$-wave,  which  plays  a  crucial  role  for  polarizability
differences. Our results for the quadrupole polarizability sums
$(\alpha_2+\beta_2)_{\pi^0,\pi^+}$, which are controlled by the
$D$-waves, are in rather good agreement
with~\cite{filkov05,filkov06}.

\vskip2mm
\noindent{\bf $\bullet$ Errors}\\
The errors quoted in tables~\Table{polartable},\Table{compa1a2b} have been
estimated as follows. The uncertainties associated with the
description of the left-hand cut has been evaluated by varying all the
input coupling constants as well as the integration cutoff $\Lambda$
which was varied between $-3$ and $-10$ GeV$^2$. Concerning
the final-state interaction, we have varied the $\pi\pi$ scattering lengths
($a_0^0=0.220\pm0.005$, $a_0^2=-0.0444\pm0.0010$)
and used, below $E=1$ GeV two different representations for the 
$I=0$ $S$-wave $\pi\pi$ phase-shift~\cite{ACGL,GKPY}. We have also
varied the resonance parameters in the $D$-wave. Concerning the errors
in the fitted parameters, the usual criterion based on increasing the
$\chi^2$ by one unit is based on the assumption  that the experimental
errors are statistical and that the correlation matrix is known. These
assumptions are not valid in the present case. We have therefore adopted a
more phenomenological criterion, considering the $\chi^2$ per point
instead of the total one and allowing it to increase by 0.5. The
errors associated with the fitted parameters and those associated with
the  input data are quoted separately in
tables~\Table{polartable},\Table{compa1a2b}. The $\pi^+$
polarizability difference $(\alpha_1-\beta_1)_{\pi^+}$
is not completely determined by the fit as it lies at the boundary of
the allowed value. No error can be quoted in this
case. Correspondingly, we quote a single error for the LEC combination
$\tilde{a}^r_1$ in table~\Table{compa1a2b}.

\begin{table}[tb]
\bc
\bt{|l|ccc|}\hline
\TT$\phantom{Fit}$\large{$\pi^0$} & $a_1^r$   & $a_2^r$ & $b^r$     \\ \hline
\TT Fit       &$-25.9\pm1.6\pm3.7$   & $8.6\pm0.8\pm1.8$ & $3.4\pm0.4\pm0.1$\\ 
Res. mod. $[$ENJL$]$ & $-39\pm4\,[-23.3]$ & $13\pm2\,[14.9]$ &
$3\pm0.5\,[1.7]$ \\ \hline 
\TT$\phantom{Fit}$\large{$\pi^+$} & $\tilde{a}_1^r$ & $\tilde{a}_2^r$
& $\tilde{b}^r$\\ \hline 
\TT Fit &$-25.0\pm2.2\ $ & $1.4\pm1.8\pm1.4$ & $0.2\pm0.3\pm0.1$ \\ 
Res. mod. $[$ENJL$]$ & $-3.2[-8.7]$ & $0.7[5.9 ]$ & $0.4[0.4]$ \\ \hline
\et
\caption{\sl Values of the three combinations of
$O(p^6)$ coupling constants at the scale $\mu=0.77$ GeV 
involved in the $\gamma\gamma\to \pi^0\pi^0$
amplitude (second and third row)  and in the $\gamma\gamma\to \pi^+\pi^-$
amplitude (fourth and fifth row). The values from a resonance model
used  in  refs.~\cite{gasserivan05,gasserivan06}  and  from  the  ENJL
model~\cite{bijprad96} are compared with the values deduced from
our fitted dispersive amplitudes. Errors are as in table~\Table{polartable}.}
\lbltab{compa1a2b}
\ec
\end{table}
\section{Conclusions}
We have reconsidered the MO dispersive representations of
photon-photon scattering amplitudes and applied them to the recent results of
the Belle collaboration~\cite{bellemori2pic_1,bellemori2pic_2,
belleuehara2pi0_1,belleuehara2pi0_2}. 
This method is general and follows
from the non-perturbative features of QCD. Its range of applicability can be
extended up to slightly above 1  GeV by taking into account the main
source of inelasticity in  $\pi\pi$ scattering. Our description of the
left-hand cut  includes the contributions of  the vector, axial-vector
as well as tensor resonances  i.e. all relevant resonances with masses
up to 1.3 GeV. We found that the tensor resonances play a
significant role in the left-hand  cut. We employ MO representations
somewhat different from previous works, keeping the left-cut parts 
in  the form  of  spectral integrals.  In  this  manner, one  avoids
polynomial ambiguities  associated with propagators  of particles with
non-zero spin. Furthermore, the spectral function displays in a clear
way  how  a  regularization  occurs  in  exchanges  of  resonances  of
different types.

We have also argued that chiral
constraints can be imposed on the subtraction constants using model
independent information on $p^6$  chiral couplings. We have shown that
a sum rule on a single $p^6$ parameter for $\pi^0\pi^0$ provides a non
trivial constraint. Making use of all
these theoretical constraints, one obtains an amplitude containing only
six parameters.  We introduce  a  cutoff on  the left-cut  spectral
integration, accounting  for a  Regge-type regularization, but  no other
cutoff. In contrast to ref.~\cite{zheng09},  the QED Born term is used
unmodified by form-factors.

We have fitted the subtraction parameters to a data set
containing 541 points on $\gamma\gamma\to\pi^0\pi^0$ 
and 1242 points on $\gamma\gamma\to\pi^+\pi^-$. 
This set is largely dominated by Belle's results.
We find good compatibility between Belle data on $\pi^0\pi^0$ and
the lower energy data from the Crystal Ball collaboration. In the case
of $\pi^+\pi^-$ we find compatibility  between the data from Belle and
the data  from Cello as well as  the data from MarkII  below $0.4$ GeV.
The compatibility  with the MarkII data  in the range  $0.4-1.0$ GeV is
more marginal. 

In the region of the  $K\bar{K}$ threshold and the $f_0(980)$ peak, we
find reasonable agreement between Belle's data and our parametrization
which assumes  a smooth isospin  limit. The charged  channel, however,
does not rule out the  possibility of some isospin violating effects in
the shape of the peak. The cross-sections around the peak in both
the neutral and charged channels are best reproduced for values of the
threshold phase $\delta(2m_K)\simeq 200^\circ\pm20^\circ$. 
From   our   calculation  we   also  obtain  the   amplitude
$\gamma\gamma\to K\bar{K}$ for $I=J=0$.  It is unfortunately
difficult to probe this part against the experimental data on
$\gamma\gamma\to K^+\bar{K}^-, K^0\bar{K}^0$ since the experimental
amplitudes contain admixtures from isospin $I=1$. 

We have shown that Belle's data are compatible with
the pion polarizabilities predicted in ChPT. In the case of the
$\pi^0$ we have derived a refined value. We have also derived results
for the quadrupole polarizabilities of both the charged and neutral
pion. Our results for the differences 
$\alpha_2-\beta_2$ are somewhat smaller than those derived in
refs.~\cite{filkov05,filkov06}. We believe this to be due to our using
more precise $\pi\pi$ phase-shifts at low energies for the $S$-waves.  
It is clear, however, that experimental data on photon-photon
scattering at low energies (i.e. below 0.5 GeV) are most efficient for
determining pion polarizabilities with precision. It is hoped that
such experiments will be performed at facilities like KLOE2 or
BESIII. 

\medskip
\noindent{\large\bf Acknowledgements:}
The authors would like to thank the Belle collaboration
and especially prof. S. Uehara for sending us their experimental
results and for correspondence.  We also thank prof. B.  Ananthanarayan for
carefully reading the manuscript and for comments.
This work is supported in part by the
EU Contract No. MRTN-CT-2006-035482 FLAVIAnet and by the EU project
HadronPhysics2.


\end{document}